\documentclass[final,5p,times,twocolumn]{elsarticle} 

\usepackage[colorlinks=true]{hyperref}

\usepackage{epsfig}

\usepackage{amsmath}
\usepackage{amssymb}

\journal{Nuclear Instruments and Methods in Physics Research A}

\usepackage{subcaption}
\usepackage{graphicx}
\usepackage{booktabs}

\newcommand{\labfig}[1]{\label{fig:#1}}
\newcommand{\reffig}[1]{\hyperref[fig:#1]{Figure}~\ref{fig:#1}\xspace}

\newcommand{\labtab}[1]{\label{tab:#1}}
\newcommand{\reftab}[1]{\hyperref[tab:#1]{Table}~\ref{tab:#1}\xspace}

\newcommand{\labsec}[1]{\label{sec:#1}}
\newcommand{\refsec}[1]{\hyperref[sec:#1]{Section}~\ref{sec:#1}\xspace}

\newcommand{\labapp}[1]{\label{sec:#1}}
\newcommand{\refapp}[1]{\hyperref[sec:#1]~\ref{sec:#1}\xspace}

\newcommand{\pat}[1]{{{#1}}} %
\newcommand{\eric}[1]{{{#1}}}
\newcommand{\fran}[1]{{{#1}}}
\newcommand{\jonat}[1]{{{#1}}}
\newcommand{\vinc}[1]{{{#1}}}
\newcommand{\sapoob}[1]{{{#1}}}
\newcommand{\saporw}[1]{{{#1}}}

\newcommand{\nimaA}[1]{#1}
\newcommand{\nimaB}[1]{#1}

\newcommand{\nimaAres}[1]{#1}

\newcommand{\nectar}[1]{NECTAr3}

\begin{document}

\begin{frontmatter}




\title{Characterization and \nimaA{performance} of an upgraded front-end-board for \nimaA{NectarCAM}}




\author[IRFU]{F.~Bradascio}
\author[IRFU]{F.~Brun}
\author[LPNHE]{F.~Cangemi\fnref{nowatapc}}
\author[LAPP]{S.~Caroff}
\author[IRFU]{E.~Delagnes}
\author[UB]{D.~Gascon}
\author[IRFU]{J.-F.~Glicenstein}
\author[CPPM]{D.~Hoffmann}
\author[IRAP]{P.~Jean}
\author[LPNHE]{C.~Juramy-Gilles}
\author[LPNHE]{J.-P.~Lenain}
\author[IRFU]{V.~Marandon}
\author[LPNHE]{J.-L.~Meunier}
\author[LPNHE]{E.~Pierre}
\author[APC]{M.~Punch}
\author[UB]{A.~Sanuy} 
\author[IRFU]{P.~Sizun}
\author[LPNHE]{F.~Toussenel}
\author[IRFU]{B.~Vallage}
\author[LPNHE]{V.~Voisin}


\fntext[IRFU]{IRFU, CEA, Université Paris-Saclay, Bât 141, 91191 Gif-sur-Yvette, France}
\fntext[LPNHE]{Sorbonne Universit\'e, CNRS/IN2P3, Laboratoire de Physique Nucl\'eaire et de Hautes Energies, LPNHE, 4 place Jussieu, 75005 Paris, France}
\fntext[LAPP]{Univ. Savoie Mont Blanc, CNRS, Laboratoire d'Annecy de Physique des Particules - IN2P3, 74000 Annecy, France}

\fntext[UB]{Departament de F\'isica Qu\`antica i Astrof\'isica, Institut de Ci\`encies del Cosmos, Universitat de Barcelona, IEEC-UB, Mart\'i i Franqu\`es, 1, 08028, Barcelona, Spain}
\fntext[CPPM]{Aix-Marseille Université, CNRS/IN2P3, CPPM, 163 Avenue de Luminy, 13288 Marseille cedex 09, France}
\fntext[IRAP]{Institut de Recherche en Astrophysique et Planétologie, CNRS-INSU, Université Paul Sabatier, 9 avenue Colonel Roche, BP 44346, 31028 Toulouse Cedex 4, France}
\fntext[APC]{Universit\'e de Paris, CNRS, Astroparticule et Cosmologie, F-75013 Paris, France}
\fntext[nowatapc]{Now at Universit\'e de Paris, CNRS, Astroparticule et Cosmologie, F-75013 Paris, France}

\begin{abstract}
This paper presents an analysis of the updated version of the Front-End Board (FEB) for the NectarCAM camera, developed for the Cherenkov Telescope Array Observatory (CTAO). The FEB is a critical component responsible for reading and converting signals from the camera's photo-multiplier tubes into digital data and generating module-level trigger signals. This study provides an overview of the design and performance of the new FEB version, including the use of an improved \nectar{}  chip with advanced features. 
The \nectar{} chip contains a switched capacitor array for sampling signals at 1 GHz and a 12-bit analog-to-digital converter (ADC) for digitization upon receiving a trigger signal. 
The integration of the new \nectar{} chip results in a significant reduction of NectarCAM's deadtime by an order of magnitude compared to the previous version. The paper also presents the results of laboratory testing, including measurements of timing performance, linearity, dynamic range, and deadtime, to characterize the new FEB's performance. 
\end{abstract}


\end{frontmatter}
%

\section{Introduction}
The Cherenkov Telescope Array (CTA) is  \jonat{the next-generation gamma-ray observatory.  The telescopes of the array are designed} to detect very high-energy (VHE) gamma-ray photons ranging from 20 GeV to 300 TeV \cite{CTAconcept}. VHE gamma rays are indirectly detected by analyzing the Cherenkov light produced in particle cascades (showers) generated when the gamma rays interact with the atmosphere. Images of the showers captured by one or multiple telescopes are analyzed to suppress cosmic-ray background and reconstruct primary gamma-ray parameters.  
The CTA observatory will comprise several tens of telescopes distributed across two sites in La Palma (Spain) and Paranal (Chile), including small-sized telescopes (SSTs), medium-sized telescopes (MSTs), and large-sized \nimaB{telescopes} (LSTs) \cite{cta_mc_design}. These telescopes are equipped with tessellated mirrors that focus Cherenkov photons onto fast-recording, pixelated cameras, capturing both spatial and temporal information of the short light pulses. The \sapoob{MSTs} at the CTA-North site will be equipped with the NectarCAM camera~\cite{nectarcam}.  

NectarCAM uses a modular structure, wherein a core component referred to as a ``module" comprises a Focal Plane Module (FPM) and a front-end board (FEB). The FPM consists of seven R12992-100-05 Hamamatsu photomultiplier tubes (PMTs) with a diameter of 1.5"~\cite{2021NIMPA100765413T}. These PMTs are accompanied by high voltage and pre-amplification boards (HVPA). Each FPM is interfaced to a FEB through an interface board (IB) hosting a micro-controller (\reffig{feb_synoptic}). Additionally, Winston cone light concentrators \cite{lightconcentrators} are mounted on the PMTs to enhance light collection efficiency.
Each FEB is connected to a digital trigger backplane (DTBP) which provides the interface to the clock and the power. The camera trigger is built on the set of \fran{265} \nimaA{DTBPs}. 
\nimaB{To achieve the scientific performances of CTA, NectarCAM should have a dynamic range of several thousand photo-electrons (p.e.) calibrated with a precision of a few percent, a sub-nanoseconds timing resolution and being able to trigger at a rate of at least 7~kHz~\cite{CTAconcept, 2017AIPC.1792h0009G}.}

\begin{figure*}[t]
    \centering
    \includegraphics[width=0.6\textwidth]{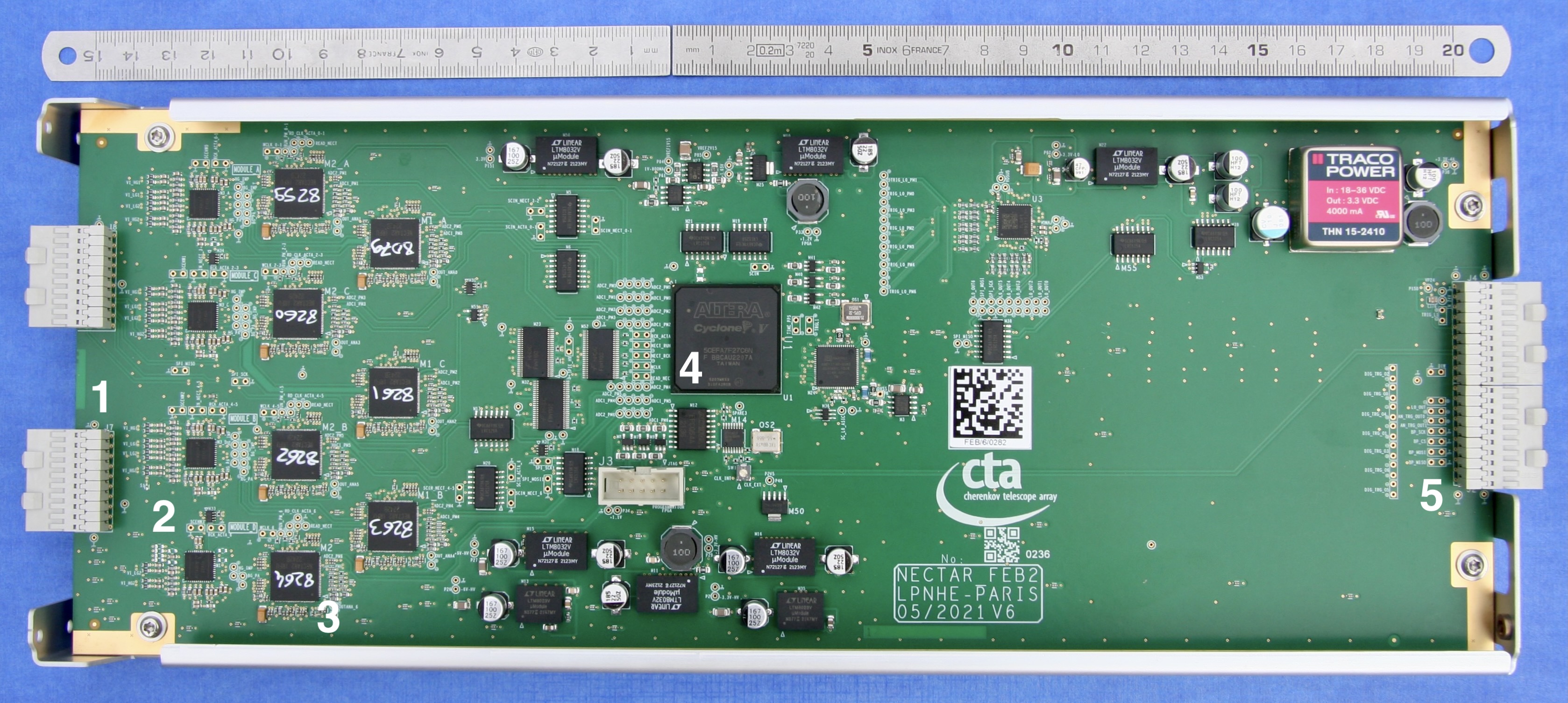}
    \caption{Front-End-Board (FEBv6) of NectarCAM. The main components are indicated by numbers. 1: Left connectors through which the electrical signals from \nimaA{7 PMTs} arrive at the board. 2: Four ACTA amplifiers. 3: Seven \nectar{} chips. 4: FPGA. 5: Right connector, which serves as the \nimaB{mechanical and electrical interface} to the DTBP. The L0 ASIC is not visible since it is located on the other side of the board.}
   \labfig{feb}
\end{figure*}

This paper provides \fran{in \refsec{FEB}} a comprehensive overview of the design and performance of the updated version of the FEB, referred to as FEBv6, shown in \reffig{feb}, which includes an enhanced variant of the \nectar{} chip. 
The integration of this advanced \nectar{} chip results in a significant reduction in NectarCAM's deadtime by an order of magnitude. {Measurements of deadtime, linearity, charge resolution, and timing precision are presented in \refsec{performances}.



\section{The front-end board}\labsec{FEB}

When Cherenkov light reaches the camera, it undergoes conversion into an electrical signal in the PMTs and is then preamplified on the HVPA board to produce two separate signals: high gain and low gain. \nimaA{The HVPA board includes a Cockroft-Walton generator as well as an amplifier, the PACTA \cite{2012JInst...7C1100S}. The low and high gain} signals are then sent to the FEB. 
The main function of the FEB is to sample, digitize and transmit these signals to the data acquisition system using the \nimaB{user datagram protocol (UDP) over Ethernet}. It also implements a local, pixel level trigger called  level 0 (L0), and the remote control of several components via serial peripheral interface (SPI)\footnote{\url{https://onlinedocs.microchip.com/pr/GUID-835917AF-E521-4046-AD59-DCB458EB8466-en-US-1/index.html?GUID-E4682943-46B9-4A20-A62C-33E8FD3343A3}}.


The FEBv6 is based on a 12-layer printed-circuit-board (PCB) with controlled impedance. A commercial supplier will produce {3000 FEBs} for the NectarCAMs of CTA. 
The main components of the FEBv6 are presented in \reffig{feb_synoptic}. The FEB {includes} 
a \fran{Cyclone V}\footnote{https://www.intel.com/content/www/us/en/products/details/fpga/cyclone/v.html} INTEL field-programmable gate array (FPGA),
described in \refsec{fpga}, which allows processing the data readout and the slow control \nimaB{data}, and three different application-specific integrated circuits (ASIC).
These ASICs are:
\begin{itemize} 
\item the L0 trigger ASIC, whose design and operations are described in \cite{L0,timing_paper};
\item four amplifiers for CTA (ACTA), which include a {basic pulse generator} for functional tests~\cite{acta};
\item seven new \nectar{} ASICs which sample and digitize the signals, described in \refsec{nectar}.
{
\eric{Due to the unavailability of the QFP128 package previously utilized, the \nectar{} chips had to be encapsulated in QFN100 packages, known for their smaller footprints. This shift in packaging is one of the main drivers behind the redesign of the FEB boards.}}

\end{itemize}
{The ACTA amplifies the low-gain and high-gain signals and sends them to the \nectar{} chip; a second copy of the high-gain signal is sent to the L0 trigger ASIC.} 
On the other \nimaAres{hand, a single copy of the low-gain signal} is solely amplified to enter a different channel of the \nectar{} chip.

The FEB runs at 66.667~MHz with either a common clock provided externally and obtained from the {DTBP,} or in single mode with an internal quartz oscillator. The implementation of a clock buffer allows switching between the two clocking options with a \nimaAres{hardware switch}.

The FEB is controlled externally by the NectarCAM module controller (NMC). The NMC is in charge of three main tasks:
\begin{itemize}
\item the dynamic and real-time configuration and monitoring of all FEBs through \pat{open platform communications unified architecture} (OPCUA\footnote{https://opcfoundation.org/about/opc-technologies/opc-ua/}) clients. NMC acts as a gateway between the OPCUA protocol from various clients to ``low-level" UDP frames understandable by the FEBs;
\item the initialization of all {FEB} configurations before the beginning of data acquisition {from} a single XML file;
\item the upload of new {firmware} to upgrade the \pat{module components  (i.e. FEB FPGA, DTBP FPGA and FPM interface board micro-controller).}
\end{itemize}




\begin{figure*}
    \centering
    \includegraphics[width=0.6\textwidth]{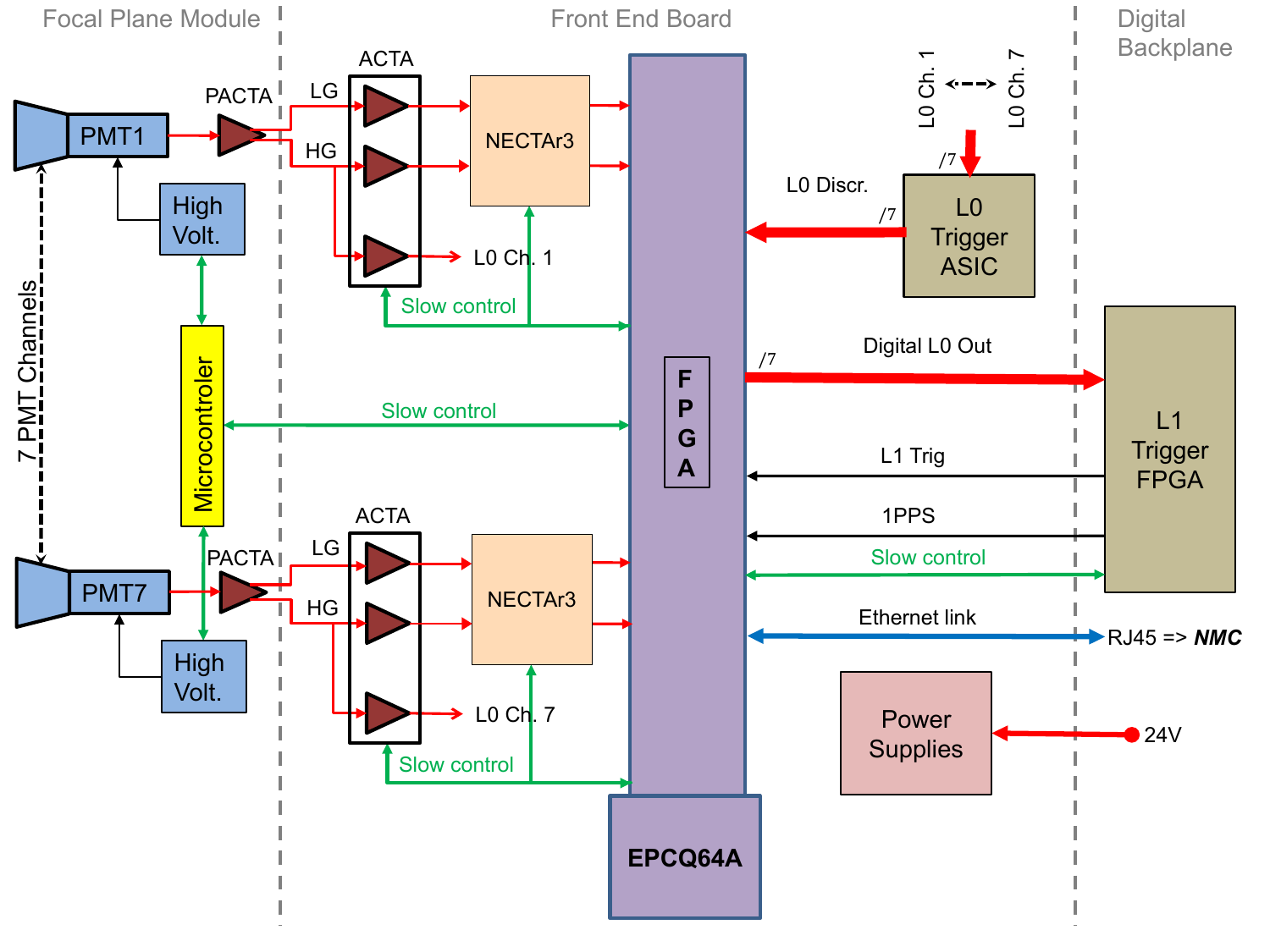}
    \caption{Diagram showing the components of the FEBv6.}
    \labfig{feb_synoptic}
\end{figure*}

\begin{figure*}[h]
    \centering
    \includegraphics[width=0.6\textwidth]{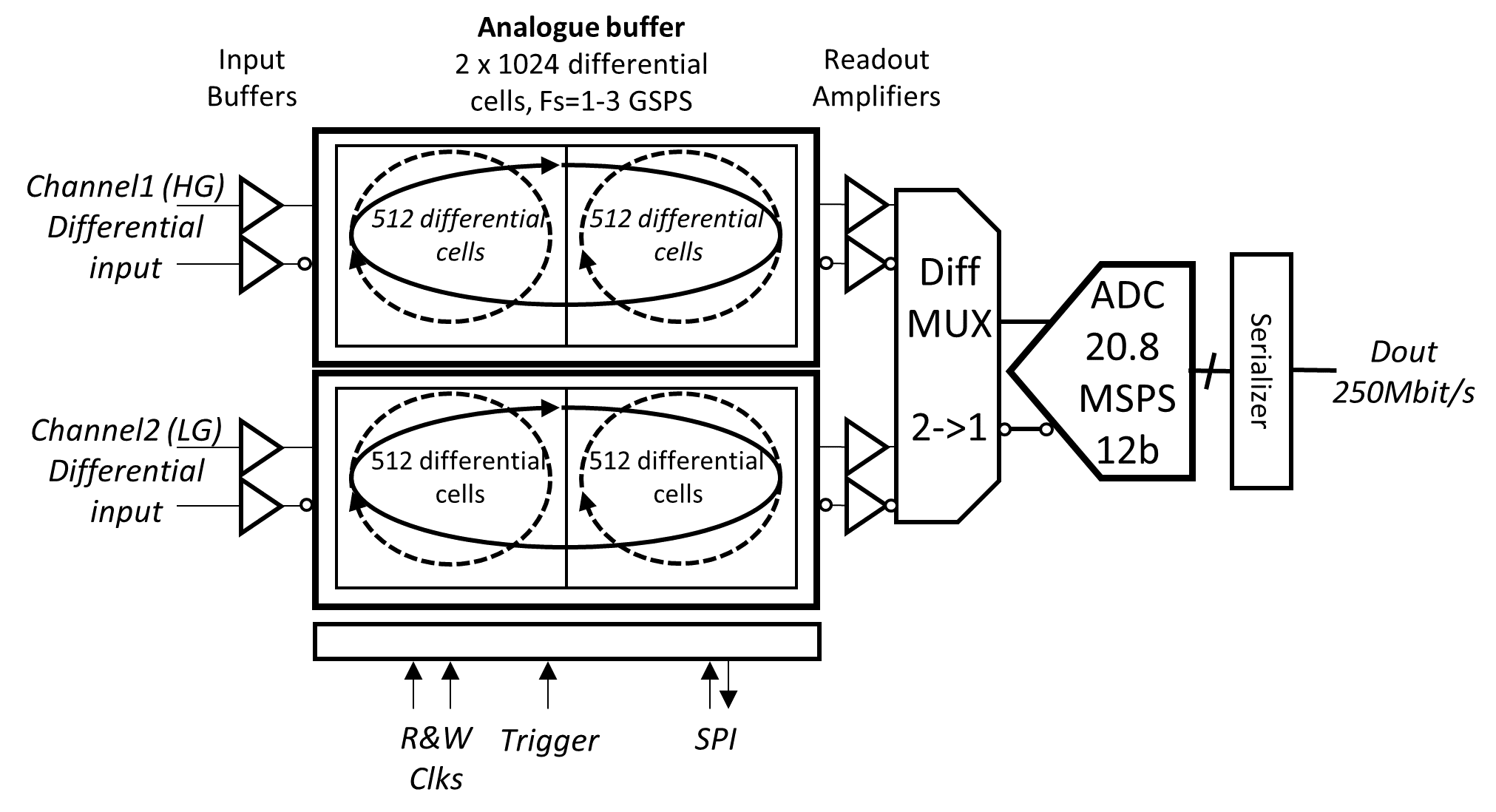}
    \caption{Block \nimaB{diagram} of the \nectar{} chips. The plain arrow corresponds to the standard write operation for \nectar{} while the dashed ones are for \nectar{} in ping-pong mode.}
    \labfig{nectarblockdiagram}
\end{figure*}

\subsection{\nectar{} ASIC}\labsec{nectar}


The \nectar{} chip's structure is illustrated in \reffig{nectarblockdiagram}.
This chip continuously samples two analog differential signals {(low and high gain)} at frequencies of up to 3 Giga samples per second (GSPS)\footnote{The sampling frequency is set to 1~GSPS for NectarCAM {(1~ns sample)}.}, employing a switched capacitor-based analog memory that functions as a circular buffer until a trigger event occurs. 
The sampling is then stopped and the samples corresponding to the zone of interest are read back and time-multiplexed by a common on-chip ADC, which digitizes them at a rate of 20.8 Mega samples per second (MSPS) before transferring them to the FPGA. The sampling is then restarted. The readout of a 60-sample window for the 2 channels typically takes 7~$\mu$s, during which the chip \nimaAres{is in a fixed deadtime}. For an input rate of 7 kHz, this slightly {exceeds a} 5\% deadtime fraction. 


The minimum depth of the {analog} buffer integrated into the chip must be sufficient to accommodate both the trigger latency and the duration of the window of interest for the event. 
In the case of NectarCAM, this requirement is less than 300~ns, a fraction of the 1024~ns depth available for 1 GSPS sampling, provided by the 1024 cells allocated for each channel of the \nectar{} chip. Taking this into consideration, we have designed the \nectar{} chip as an enhanced version of NECTAr, utilizing the AMS 350~nm CMOS technology as in previous iterations. The NECTAr chip has been already utilized in the upgraded cameras of the H.E.S.S. experiment~\cite{HESSUpgrade2019} and is extensively described in \cite{Nectar2012, nectar0}.

In the \nectar{} chip, the {analog} memory can be optionally divided into two sub-arrays of 512 cells, allowing the signal to be sampled alternately, to reduce readout deadtime.
The signal is continuously written into the first sub-array, which functions as a circular array. Upon triggering, this sub-array is frozen and digitized, while the incoming signal is simultaneously written into the second sub-array. Subsequently, at the next trigger event, this second sub-array is then frozen and read, while the write operation switches back to the first sub-array, and so on. This ``ping-pong" operation, facilitated by simultaneous read and write access, significantly reduces the readout deadtime of the \nectar{} chip, limiting \nimaAres{deadtime} to two components. The first component is the constant deadtime ($DT1$) required to fill the new memory when the write operation switches from one memory to the other. The second component, $DT2$, is not constant and occurs when a new trigger event occurs while events are being read in both sub-arrays.
In this case, it becomes necessary to wait for the completion of the read operation before initiating any memory write process. To ensure camera-wide synchronization, both of these deadtimes are controlled at the camera level by inhibiting the trigger source, and they apply uniformly to all pixels. As detailed in \refsec{deadtime}, with a {design trigger rate} of 7~kHz, the newly implemented ping-pong mode reduces the overall camera deadtime fraction by a factor of {10.4 down to $0.5\%$}.


\begin{figure*}
    \centering
    \includegraphics[width=0.6\textwidth]{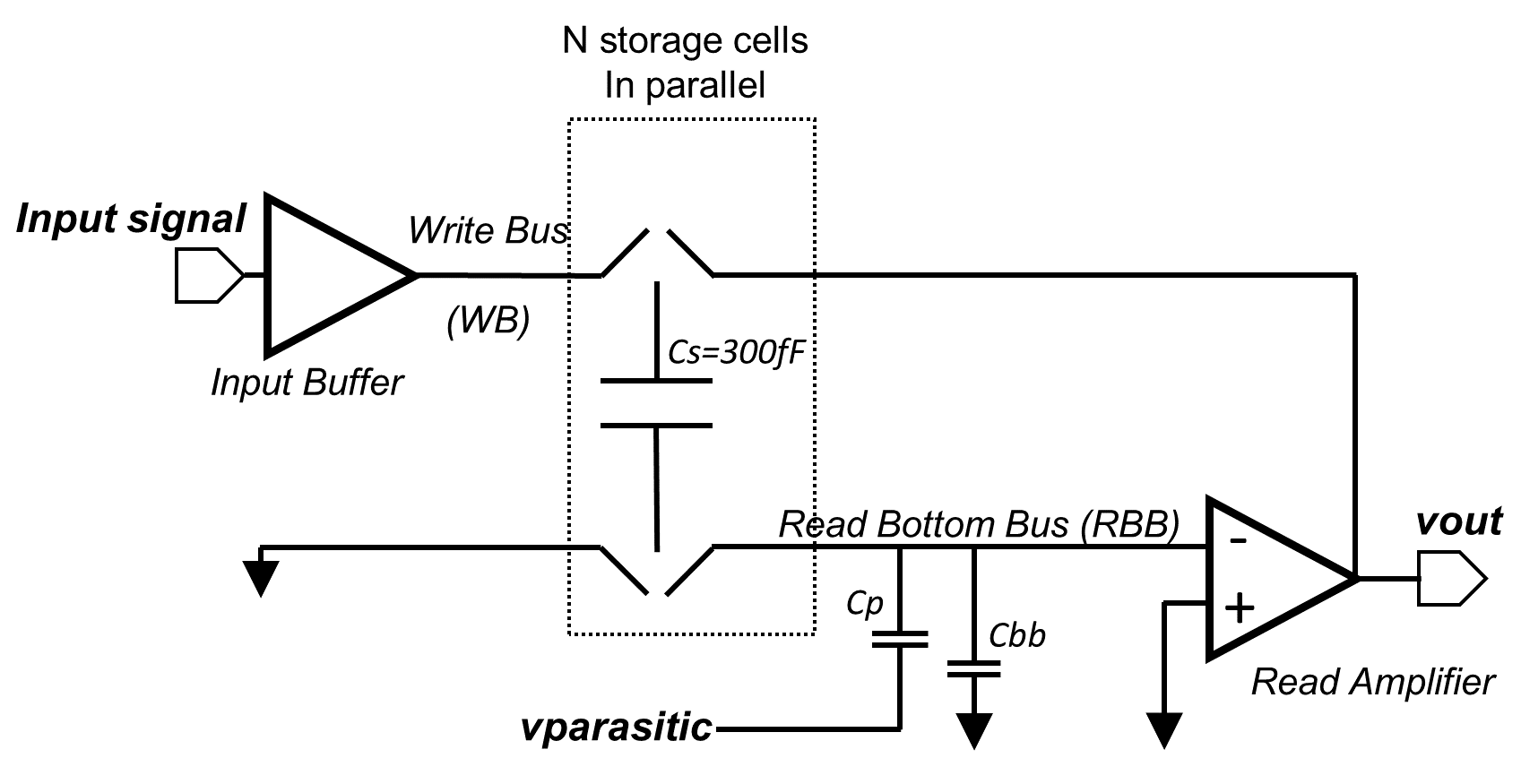}
    \caption{Main elements used for the analog write and read operations in an elementary memory channel of NECTAr (the down arrows correspond to ground).}
    \labfig{nectarRWscheme}
\end{figure*}
\begin{figure*}
    \centering
    \includegraphics[width=0.8\textwidth]{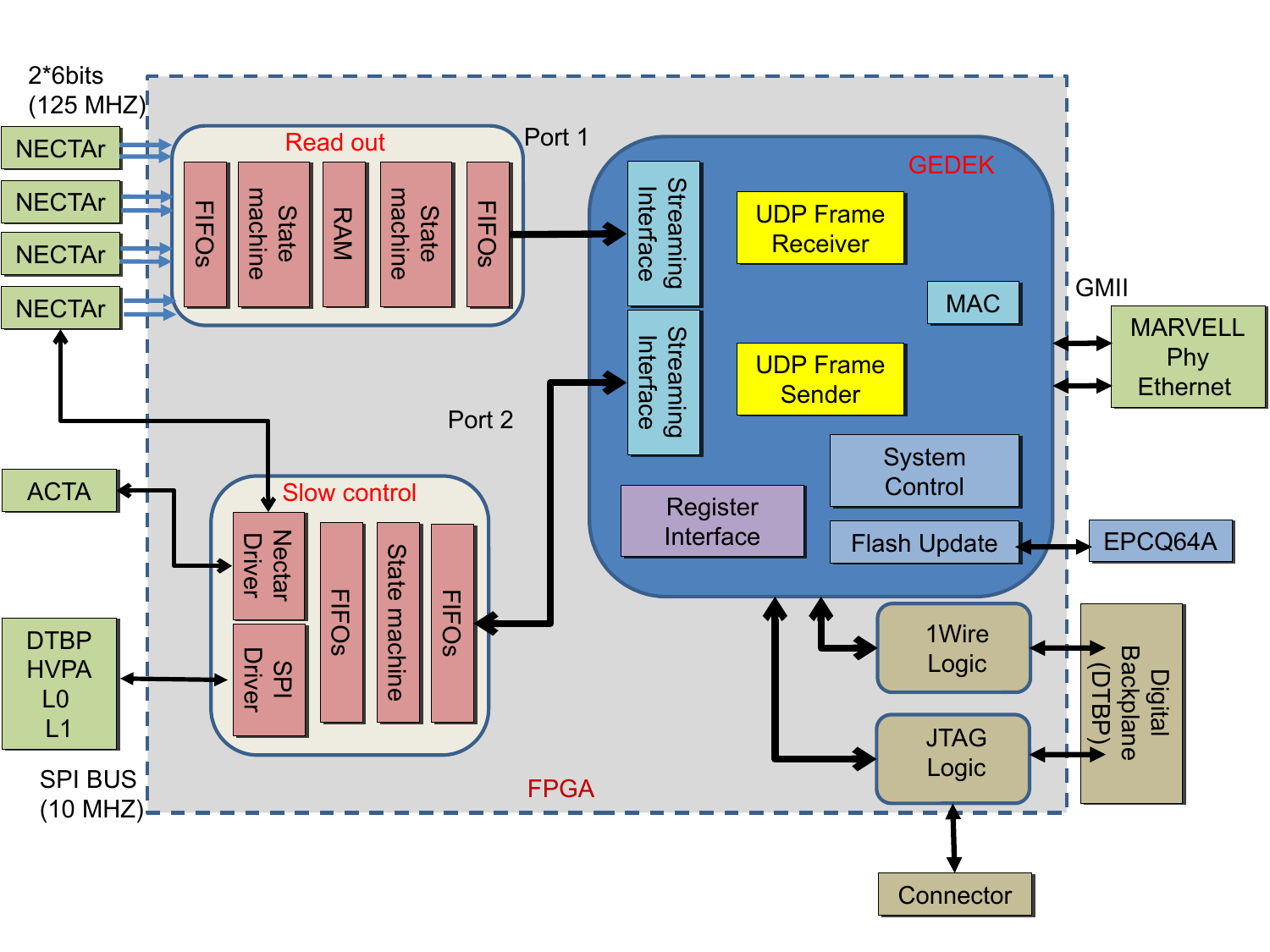}
    \caption{Operations of the FPGA on the FEB. The FPGA runs three main processes: the GEDEK Ethernet Intellectual Property (right), a control and configuration process (bottom left) and the {readout} process (top left). \sapoob{The DTBP, HVPA, ACTA, \nectar{} and L0 components are introduced in section \refsec{FEB}. The L1 (camera) trigger is created in the DTBP. The slow control part of the FPGA communicates with these components through a SPI bus. The firmware of the components is loaded through serial or JTAG protocols. Upon trigger, the NECTAr waveforms are {read out to} a RAM located in the readout part, then sent to a {control-room based camera server} through the GEDEK interface, using the UDP protocol. {The GMII interface connects the FPGA MAC to a MARVELL
    Ethernet transceiver.} The {EPCQ64A} PROM stores configuration data.}} 
    \labfig{firmware}
\end{figure*}
In ping-pong mode, the \nectar{} chip allows concurrent writing and reading operations in the analog memory. {The HG and LG channels are both differential. Each uses two complementary signals of opposite polarity. Each of these four signals is processed in the NECTAr chip using identical structures, called elementary memory channels, illustrated in \reffig{nectarRWscheme}.} A critical configuration for the system is the coupling between a large (up to 2~V) {analog} signal written on the high gain (HG) channel and a simultaneous signal read-back on the low gain (LG) channel. Although the two lines involved (HG write bus and LG read bottom bus (RBB)) are relatively far apart and  separated by other metal lines, since they run parallel over a long distance (4~mm), they are still coupled by a parasitic capacitance 
that is not properly {removed} by standard {design tools}. To optimize the layout of the line (spacing, stacking, and width) and the shielding in \nectar{}, we used 2D finite element modeling of the entire technological stack around the RBB area using the \texttt{FEMM} software~\cite{FEMM}. 
Without changing the value of Cbb (\fran{the value of the capacitance of the readout line to the ground}, see \reffig{nectarRWscheme}) this {reduced} 
the parasitic capacitance from 2~fF to 0.25~fF, ensuring that the measured HG to LG coupling effect is smaller than one least significant bit (LSB).

When reading triggered events, each sample's data from the two channels of the {analog} buffer is multiplexed alternately towards a pipeline ADC, as described in \cite{NECTARADC2010}. This ADC integrates ten cascaded 1.5-bit multiplying digital-to-analog converter (MDAC) stages, followed by a 2-bit analog-to-digital converter (ADC) final stage. As there is no voltage limiter inside NECTAr, signals with a voltage exceeding the ADC's coding range of 2V can arrive at the ADC input. As explained in \refapp{saturat}, such saturation of the ADC can degrade the quality of the LG data, \vinc{requiring special care to mitigate this effect. }

The operation and control of the \nectar{} chips and other ASICs on the FEB are managed by the FPGA. Further details about the board's management are provided in the next section.

\subsection{Management of the FEB}\labsec{fpga}

The board management is performed by a Cyclone V FPGA, {which is programmed with} different processes based on finite state machines (FSMs), random access memories (RAMs) for event storage (calculating charges and samples of each channel), and first-in-first-out (FIFO) buffering. These processes are synchronized by distributed clocks (e.g. 66~MHz, 125~MHz) generated from the {main 66.67~MHz clock} via a phase-locked loop.

As shown in \reffig{firmware}, three main processes are implemented in the FPGA:
\begin{itemize}
\item The first one (shown in the right side of \reffig{firmware}) is an Ethernet Intellectual Property (GEDEK\footnote{https://www.alse-fr.com/sites/alse-fr.com/IMG/pdf/gedek\_intro.pdf}) {interfacing between the FEB and the NMC} via UDP at 1~GB/s.
\item The second process (bottom left of \reffig{firmware}) handles the configuration of all ASICs within the FEB module via SPI at 10~MHz. Each component of the FEB module has its own SPI timing characteristics. To control them, the FPGA implements a SPI generic driver, which is adaptable in terms of data width and timing specifications. 

\item The third process (located at the top left of \reffig{firmware}) is responsible for controlling the readout in camera acquisition mode. The camera acquisition mode includes calibration data, such as pedestals or gains, as well as sky data. 
The data-taking can be configured to obtain time-stamps, charges, and {$N_{f}$} samples\footnote{{$N_{f}$} is the number of samples acquired per pixel for every triggered event.} into one observation window  (with {$N_{f}$} ranging from 1 to 60 at a sampling frequency of 1~GSPS). The process {reads out} the \nectar{} ASIC {as} 12-bit samples on a 2-bit serial bus at 125~MHz, with specific synchronization after triggering. Furthermore, it formats data (charges and/or samples) from all channels of one event into one frame of 32-bit words.
\end{itemize}



The FEB {has} only one Ethernet physical port (PHY from Marvell\footnote{https://www.marvell.com/products/ethernet-phys.html}). {To support both virtual data and slow-control ports, the GEDEK Intellectual Property manages dynamic IP addressing.}
This allows for data acquisition as well as monitoring different parameters of the module, such as temperature, and high voltage of PMTs. Most FPGA parameters, such as the {$N_{f}$} value, SPI timing characteristics, and ACTA's pulser frequency, are remotely configurable via UDP on the slow-control port. {Configuration data of the FPGA are stored on an EPCQ64A EPROM.}

Regarding the identification of front-end modules, the FPGA also implements a 1-Wire\footnote{https://www.analog.com/en/technical-articles/guide-to-1wire-communication.html} protocol for reading the MAC address included in a DS2502-48\footnote{https://www.analog.com/en/products/ds2502-e48.html} in the DTBP. Additionally, an initializing process {updates} the FEB with one Internet Protocol (IP) address defined by the user.

Lastly, one of the requirements is to be able to remotely program all programmable components of the front-end module. As a result, the firmware implements several processes to load the firmware of these components via UDP through serial or JTAG\footnote{https://standards.ieee.org/ieee/1149.1/1728/} protocols.

As mentioned in \refsec{FEB}, a series of functional and electronic tests are performed on each individual FEB. The next section is dedicated to the study of the FEBv6 performances in the NectarCAM.





\section{\saporw{Performance}}
\labsec{performances}
\vinc{The results presented in this section are based on tests of {ten pre-production FEBs}. They were {installed} in the first NectarCAM unit and tested in the dark room of CEA Paris-Saclay.} The dark room is temperature controlled and equipped with all camera services and calibration light sources. It is connected to a control room in which the full data acquisition (DAQ), storage and run control systems are located.
To assess the performance of the FEBs, three distinct light sources were utilized: a flat field calibration light source (FFCLS), a continuous night sky background (NSB) source, and a laser source. A detailed description of these sources can be found in \cite{timing_paper}. 

In the following sections, we {report FEB} performance evaluation, specifically deadtime, linearity, charge resolution, and single pixel timing precision.


\subsection{Deadtime}\labsec{deadtime}


In detection systems that record discrete events, deadtime refers to the duration after each event during which the system is unable to record another event. In the case of NectarCAM, the total deadtime primarily arises from two factors: the readout process of the \nectar{} chip and the buffering within the data acquisition system.

The deadtime can be measured using random Poisson sources. We used two different random generators: one based on the output of a photomultiplier triggered by a photon source \cite{timing_paper}, and one using the NSB source. 
{The first} random generator triggers the FFCLS and the laser sources. The trigger rates can be adjusted by changing the intensity of the source in front of the photomultiplier.  
The second {generator} uses the NSB source, with an internal threshold set to 6~p.e., and {an} internal trigger. The trigger rate is adjusted by changing the current of the NSB source. 
For each measurement, the camera trigger rates and the busy events\footnote{An event is flagged as busy when the trigger management board cannot send triggers to the FEBs because they are digitizing an event \cite{timing_paper}.} are also measured using the camera server. 

The deadtime fraction has been evaluated in two different ways. If no busy triggers are lost, as is the case for NectarCAM, the deadtime fraction  can be estimated using the formula:
\begin{equation}
    \delta_{\mathrm{fraction}} = \frac{N_{\rm busy}}{N_{\rm collected}}
\end{equation} 
where $N_{\rm busy}$ and $N_{\rm collected}$ are the number of busy and total collected\footnote{The triggered events can still be collected even if the front-end-boards are busy.} triggers, respectively.

\begin{figure}
    \centering
    \includegraphics[width=\columnwidth]{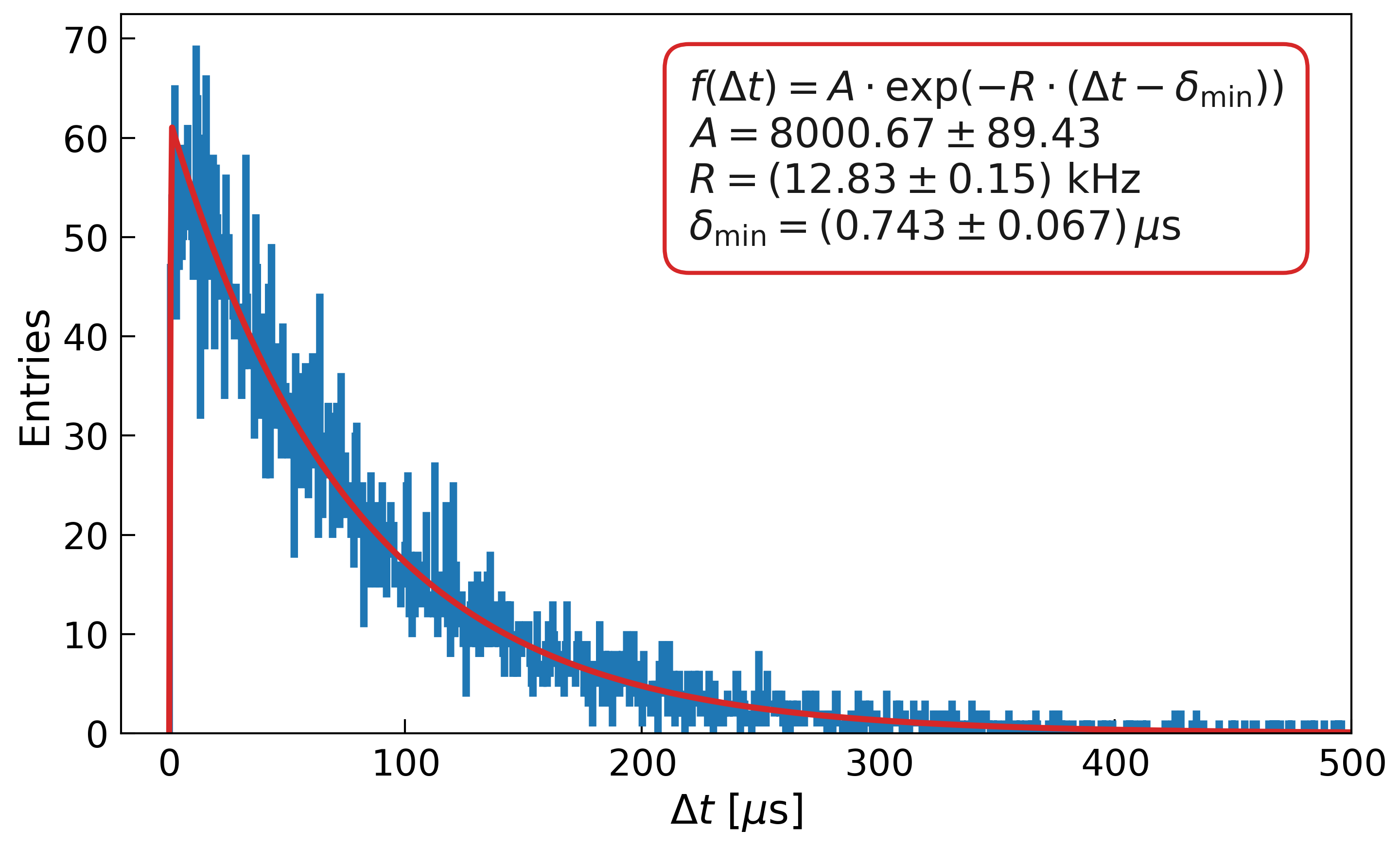}
    \caption{Distribution of time differences (in $\mu$s) between 2 triggers using the NSB source with 32~mA current. The truncated exponential fit of \autoref{eq:fit} is shown in red. The exponential parameter $R$ represents the trigger rate.}
    \labfig{expfit}
\end{figure}

The second method estimates the deadtime fraction from the distribution of the time difference $\Delta t$ between subsequent events. For events occurring randomly according to a Poisson distribution, the probability of an event occurring within a time interval $[t, t+dt]$ with no events occurring between $t=0$ and $t=\delta_{\rm{min}}$ is ideally described by an exponential distribution, where $\delta_{\rm{min}}$ represents the minimum deadtime. We fit the distribution of the time difference $\Delta t$ between subsequent events by a truncated
exponential fit \cite{GAO2022166823}:
\begin{equation}
f(\Delta t; \delta_{\rm{min}}, R)= 
\left\{ \begin{array}{l@{\;\;\;\;}r}
A \exp{\left(-R \left( \Delta t - \delta_{\rm{min}}\right)\right)}  & \Delta t> \delta_{\rm{min}} \\ 
0 & \rm{otherwise}\end{array} \right.
\label{eq:fit}
\end{equation}
where $R$ is the rate parameter of the Poisson process, and $\delta_{min}$ is the minimum deadtime of the events.
The deadtime fraction in this ideal case is obtained by multiplying $\delta_{\rm{min}}$ by the total trigger rate $R$ obtained from the truncated exponential fit.
In reality, some triggers could be lost when both arrays of the ping-pong mode of the \nectar{} chip are occupied, resulting in a non-exponential distribution. This is especially true at high rate $>15$~kHz.
However, we can still approximate the distribution of $\Delta t$ with an exponential law, as demonstrated in \reffig{expfit}. 
The figure displays an example of the distribution of time differences between successive events obtained using the NSB source with a 32~mA current corresponding to a trigger rate of $\sim13$~kHz.

\begin{figure}
    \centering
    \includegraphics[width=\columnwidth]{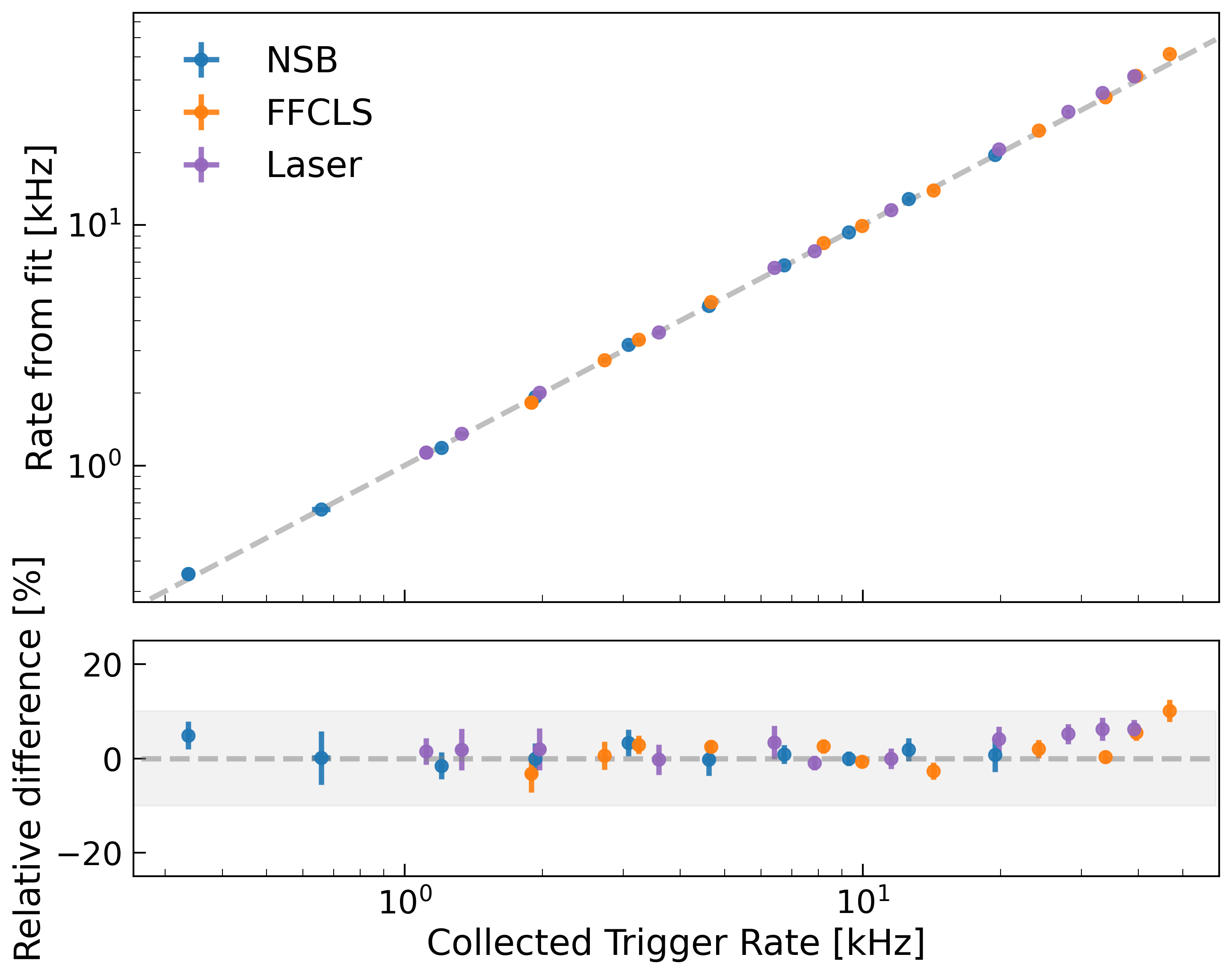}
    \caption{Trigger rate derived from the exponential fit as a function of the collected trigger rate for all runs for the measurement using the NSB source (in blue), the random generator with the FFCLS (in orange) and the random generator with the laser  (in violet). The dashed line represents the \nimaAres{perfect agreement}. The bottom plot displays the relative difference between the trigger rates obtained from the fit and the collected trigger rates as a function of the collected trigger rate. The $10\%$ band is depicted in gray.}
    \labfig{rate}
\end{figure} 

The \nimaA{truncated exponential fit} is performed for all runs and all source measurements. The exponential parameter $R$ represents the rate at which events are collected. Thus, it can be compared with the trigger rate recorded by the acquisition system. As shown in \reffig{rate}, the rate recovered from the fit is comparable to the collected rate recorded by the acquisition system for all three measurements within 10\%. Above 10~kHz, the two estimates diverge because events start to be lost in the DAQ FIFO (refer to the results of the MC simulation comparison later in this section).

\begin{table}[]
    \centering
    \resizebox{0.8\columnwidth}{!}{%
    \begin{tabular}{c ||c  }
         \toprule         \textbf{\sc{Light Source}} & \textsc{Minimum Deadtime $\delta_{\rm{min}}$} [ns]\\
         \midrule
         \sc{NSB} & $742.9\pm12.9$\\
         \sc{FFCLS} & $713.2 \pm 0.1$\\
         \sc{Laser}  & $738.1\pm 1.0$\\  
         \hline
         
         \sc{Weighted Average} &  $713.5 \pm 7.2$\\  
         \bottomrule
    \end{tabular}
    }
  
    \caption{\nimaA{Minimum deadtime ($\delta_{\rm{min}}$) values obtained from the truncated exponential fit of the $\Delta t$ distributions for various light sources, averaged over all measurements and including systematics.}}
    \labtab{deadtime}
\end{table}


\begin{figure}[h]
    \centering
    \includegraphics[width=\columnwidth]{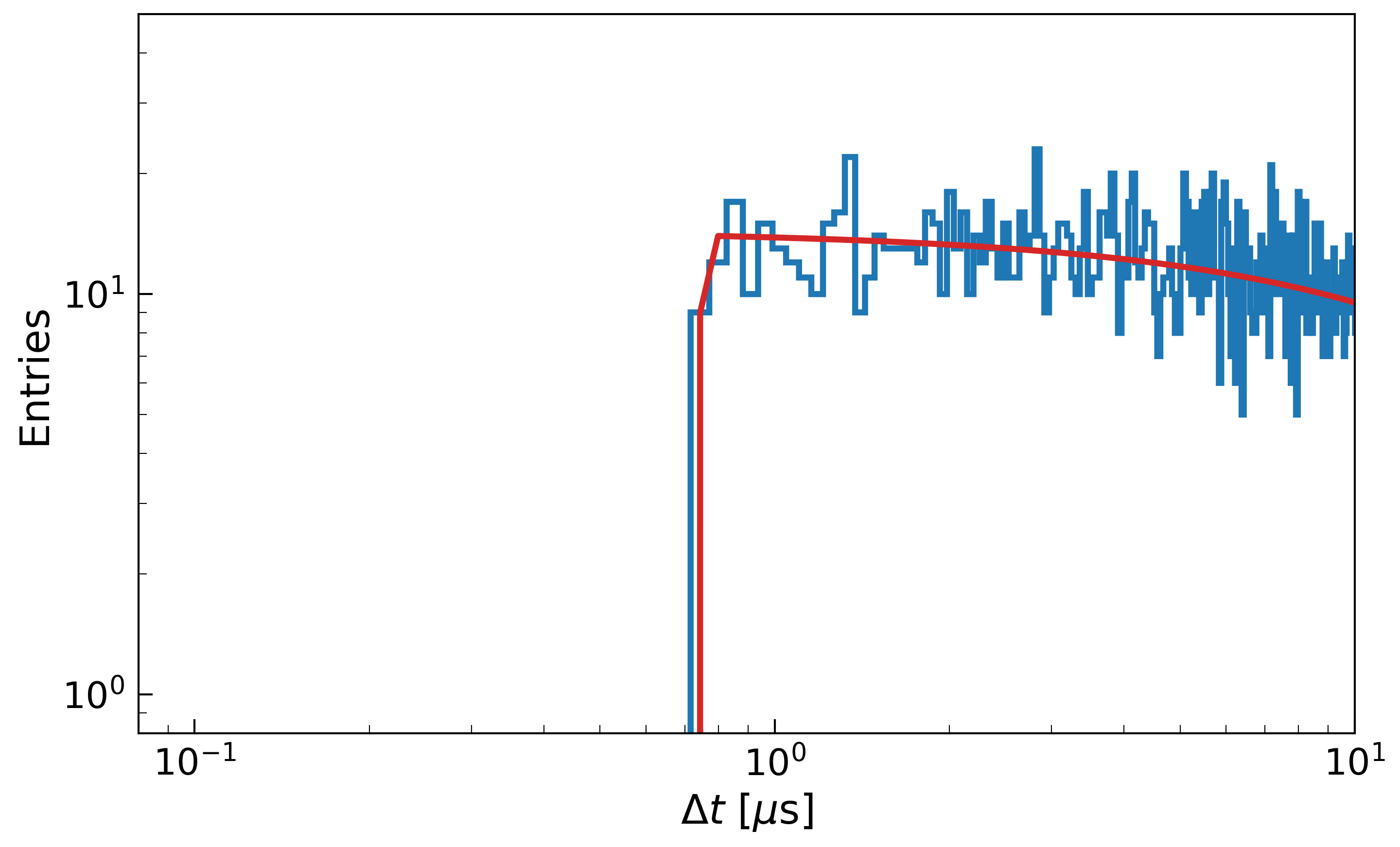}
    \caption{Distribution of time differences (in $\mu$s) between 2 triggers for time differences below $10~\mu$s taken with the laser source triggered by a random generator. The truncated exponential fit of \autoref{eq:fit} is shown in red.}
    \labfig{deadtime_histo}
\end{figure}
\reffig{deadtime_histo} illustrates the distribution of consecutive time differences for a measurement taken using the laser source, focusing on time differences below $10~\mu$s. The minimum time gap between two events, measured through the truncated exponential fit in \autoref{eq:fit}, is approximately $\delta_{\rm{min}}\simeq700$~ns across all measurements. This gap represents the minimum deadtime. \reftab{deadtime} presents the deadtime values obtained for each light source, calculated by averaging over all runs. The dispersion in deadtime values reflects the systematics in minimum deadtime estimation methods. The final minimum deadtime value is determined by taking a weighted average over all sources and yields $\delta_{\rm{min}} = 731.5 \pm 7.2$~ns, including the systematics of the sources. The error of $7.2$~ns is dominated by the spread values obtained with the three random generators. 
The deadtime of $\sim700$~ns was in fact imposed by the DAQ, so that the read-out deadtime of the \nectar{} chip is smaller than this value and is not anymore the limiting factor of the camera deadtime.

The results of the two methods of calculation of the deadtime fractions are shown in \reffig{deadtimerate}. The deadtime fraction in percentage is plotted as a function of the collected trigger rate for the three measurements using the FFCLS, the laser and the NSB sources. The first method results are shown by the shadow areas, while the second method results are illustrated with dots. As shown by the blue, orange and violet lines, at 7~kHz we have $0.5\%$ deadtime.  
The large uncertainties on the NSB source results are due to low statistics at low trigger rates, as shown by the size of the blue band.

The two methods agree up to a few tens of kHz, but they slightly differ from each other at high trigger rates. The first method is more reliable than the exponential fit, which addresses only the \nectar{}-related deadtime and thus provides only a quick cross-check of the trigger rate and minimum deadtime values.
This discrepancy is due to the loss of some busy triggers or overflows in the FIFO queue system. This behavior can be demonstrated through MC simulations. For comparison, the deadtime fraction measurements using the FEBv5 (not having ping-pong mode) with the laser source are also shown\nimaAres{,} demonstrating that at 7~kHz rate the new ping-pong mode reduces the deadtime fraction by a factor of 10.4 to $0.5\%$.

\begin{figure}[t]
    \centering
    \includegraphics[width=\columnwidth]{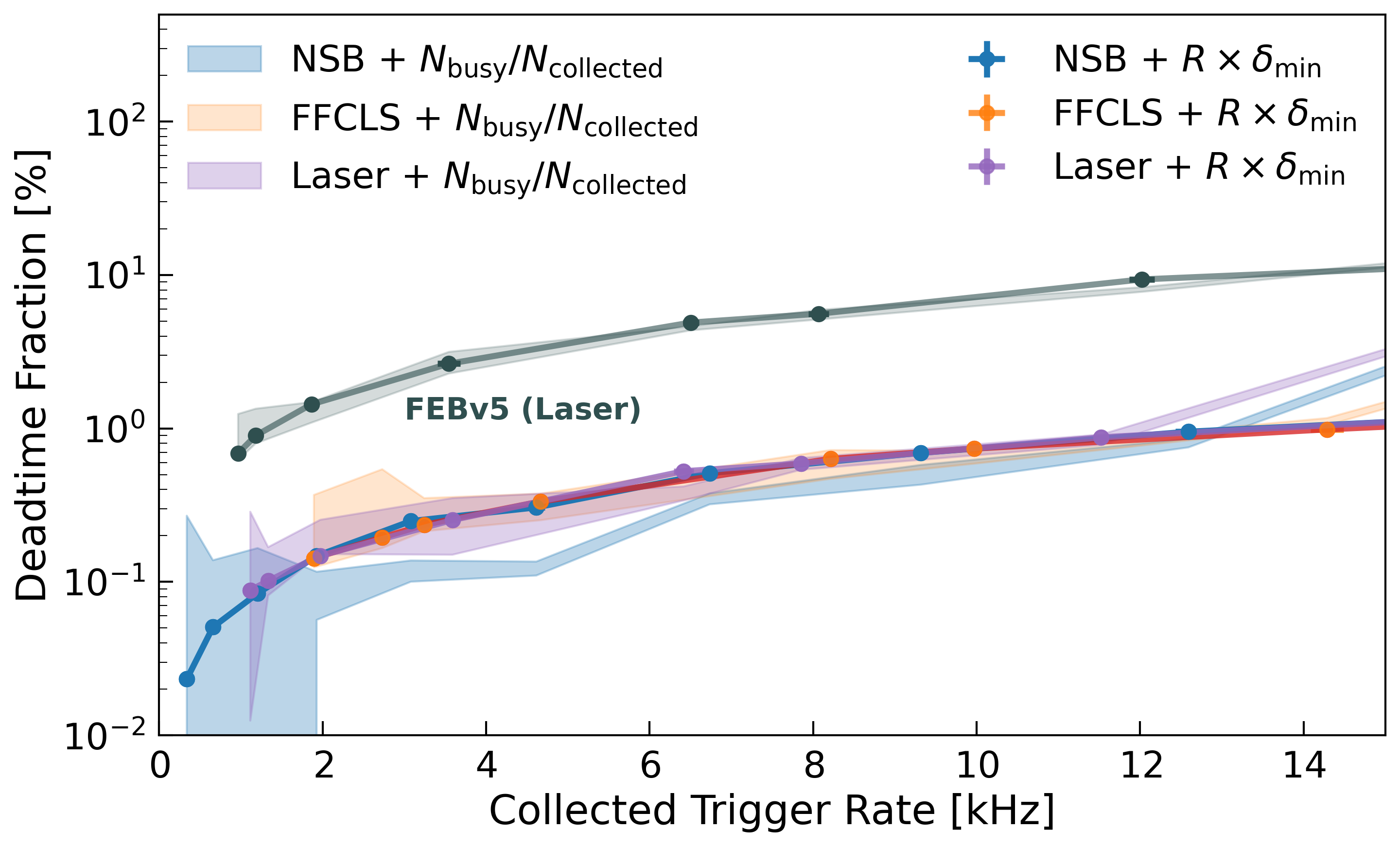}
    \caption{Deadtime fraction for the 10 FEBv6 (in color) and the FEBv5 (in gray). The deadtime fraction estimated by the ratio between the busy trigger rate and the total trigger rate (filled area) is compared with that obtained from the exponential fit of \autoref{eq:fit} (dots and solid line). All the three sources are shown: the random generator with the FFCLS (in orange), the random generator with the laser source (violet) and the NSB source (blue) measurements are shown.}
    \labfig{deadtimerate}
\end{figure}

\reffig{deadtimesim} shows the comparison between measurements and MC simulations for the FEBv5 and FEBv6. \sapoob{The figure shows that the method 2 is validated by simulations and that the deadtime is understood.} \sapoob{The simulation shows that the increase in deadtime fraction above 15~kHz is due to the filling up of the Ethernet FIFO in the FEB (see read out block in \reffig{firmware}).} In the MC simulation, the readout times for 60 samples is set to $7~\mu$s for \nectar{} and $8.6~\mu$s for \nectar{} in ping-pong mode, requiring the readout of 16 additional samples. 

\begin{figure}[h]
    \centering
    \includegraphics[width=\columnwidth]{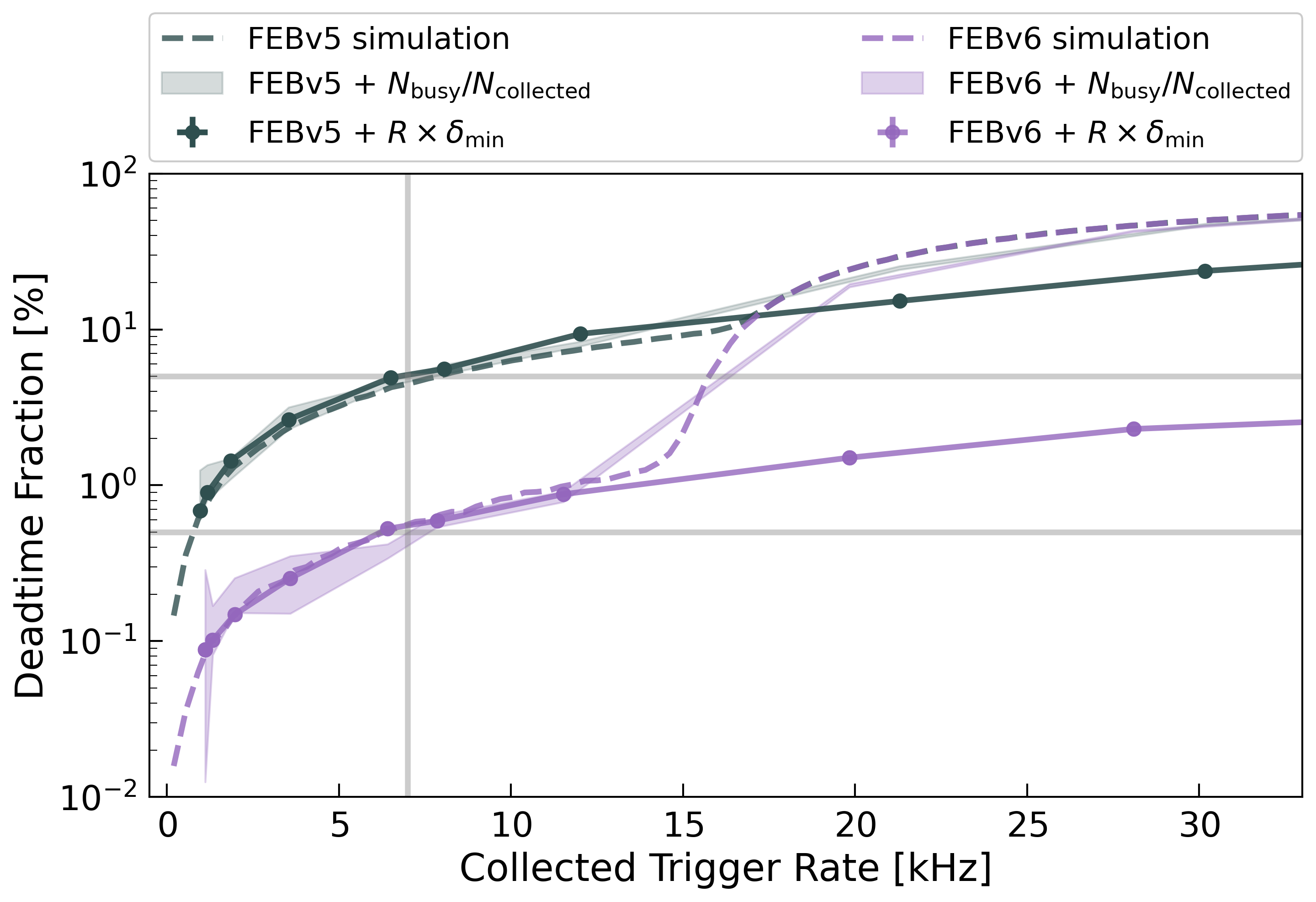}
    \caption{Comparison between measurements using the laser source and MC simulations for the FEBv5 (in gray) and FEBv6 (in blue). The method using the exponential fit and the busy rate are shown by dots  and filled area, respectively, while simulation results are shown by dashed lines. The gray lines show the deadtime fraction value at 7~kHz for the two FEB versions. }
    \labfig{deadtimesim}
\end{figure}



\subsection{Linearity}
The linearity test 
quantifies the output distortion with the increase of the incident light intensity at a given gain.  The goal is to show that the light measured by the \vinc{combined FEB and FPM} modules (``estimated" charge) is linearly proportional to the input light (``\jonat{illumination}" charge). 
The linearity of the 10 FEBs' readout has been measured by recording the charge deposited in every pixel when illuminated by the light of the FFCLS attenuated by \nimaA{one or more} \jonat{calibrated} filters.  A series of \textit{Edmund} filters\footnote{\url{https://cdn.coverstand.com/30093/556052/1e10501890efd93e0b9a5b8a644dd99d07683283.pdf}} have been used to obtain an illumination in the range 0.1--3000~p.e.
\jonat{The advantage of using calibrated filters for the linearity test is that it does not rely on the calibration of the light output of the FFCLS by an external device.}

The charge is extracted by integrating the PMT waveform in a 16~ns time window around the main peak after baseline subtraction. \pat{For} each measurement, the deposited charge is obtained by averaging over all events and pixels. \autoref{fig:charge_vs_transmission} shows the average collected charge in ADC counts as a function of the filter's \nimaA{transmission fraction}.
\jonat{A filter \nimaA{transmission fraction} of 1 corresponds to no filter.}


\begin{figure}[t]
    \centering
    \includegraphics[width=\columnwidth]{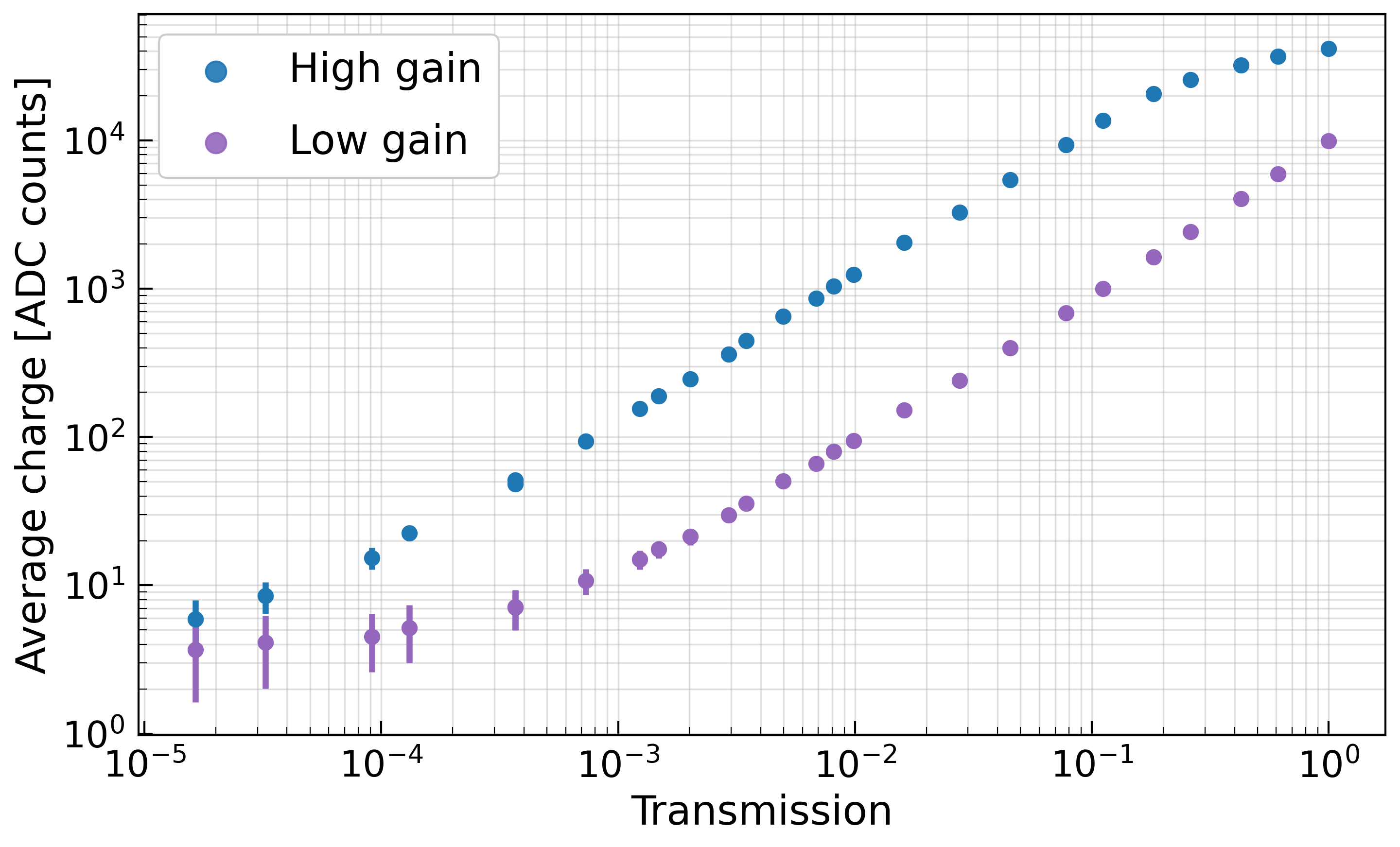}
    \caption{Average deposited charge over all pixels as a function of the \nimaA{transmission fraction} of the filters for the high gain (in blue) and low gain (in violet) channels. }
    \label{fig:charge_vs_transmission}
\end{figure}

The \jonat{illumination} charge is obtained by normalizing the output of the flasher at one intensity point  (the charge of the high and low gain channels corresponding to a transmission of \jonat{0.01}) and obtaining the other intensity points from the filter transmission. 




The output of each channel is fitted with respect to the transmission of the filters and then converted into p.e. The value of the illumination charge in p.e. is obtained by assuming the deposited charge and illumination charge are the same at transmission 0.01.
The deposited charge in \nimaB{both channels} is obtained by dividing the charge in ADC counts by a correction factor\footnote{This value is the theoretical value of 1~p.e. in ADC counts.} of 58~ADC \nimaA{counts/p.e.}. 



The results of the analysis are shown in \reffig{linearity}. 
The top panel shows the deposited mean charge as a function of the illumination charge, for the high gain channel (in blue) and the low gain channel (in violet) after applying the conversion in units of p.e. For both channels, a linear weighted least square fit has been performed. Since the high gain is linear up to $\sim400$~p.e, the linear fit has been performed only below this value. For charges larger than $\sim 400$~p.e., the saturation regime of the HG starts.  The results of the fit for the HG data are:
\begin{align}
    a_{\mathrm{HG}} & = 0.98 \pm 0.01\\
    b_{\mathrm{HG}} & = (0.09 \pm 0.03)~\mathrm{p.e.}
\end{align}
where $a_{\mathrm{HG}}$ and $b_{\mathrm{HG}}$ are the \jonat{slope} and the intercept of the linear function, respectively. The  $\chi^2$ of the fit is 2.1 for 14 degrees of freedom. 

\vinc{The linear fit of the low gain channel has been performed between $\sim 10$~p.e. up to $\sim2000$~p.e.}, where the channel is linear. The results of the fit for the LG data are:
\begin{align}
    a_{\mathrm{LG}} & = 0.97 \pm 0.01 \\
    b_{\mathrm{LG}} & = (0.93 \pm 0.31) ~\mathrm{p.e.}
\end{align}
where $a_{\mathrm{LG}}$ and $b_{\mathrm{LG}}$ are the \nimaA{slope} and the intercept of the linear function, respectively. \vinc{The $\chi^2$ of the fit is 4.3 for 13 degrees of freedom. }


The slopes of the two gain channels are compatible with 1, showing that in the corresponding fitting regions both gains are in the linearity regime. 
This is also demonstrated by the second panel in \reffig{linearity}, showing the residuals of the fit of both gain channels. The residuals are expressed as charge fraction, obtained by dividing the difference between the measured charge and the one expected from the fit by the \nimaB{expected} charge.  The linearity is better than 5\% between 0.04 and 400~p.e. for the high gain, and between 20 and \nimaA{2000~p.e.} for the low gain. \jonat{The 5\% limits are shown in the bottom panel of \reffig{linearity} by dashed gray lines.}  The overall dynamic range of the readout is therefore greater than 3 decades.
 


The high gain-low gain ratio is shown in the bottom of \reffig{linearity}. The ratio can be fitted by a constant value of 13.1 between 7~p.e. and 400~p.e. \nimaA{The range between 7~p.e. and 400~p.e. can thus be used for online cross-calibration of the HG and LG channels.}

\begin{figure}[t]
    \centering
    \includegraphics[width=\columnwidth]{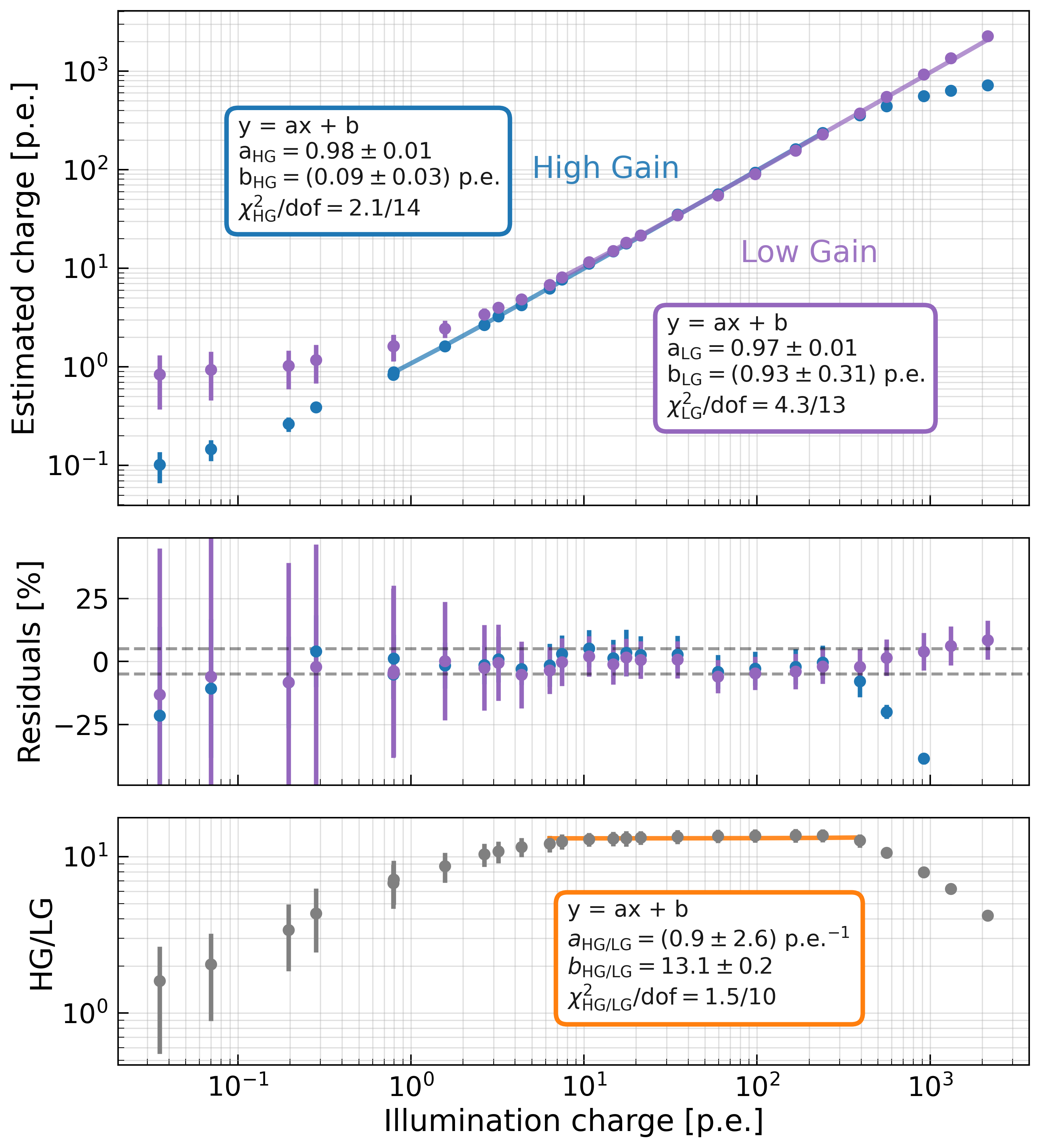}
\caption{Linearity of the FEB modules. The top frame shows the measured charge as a function of the input pulse intensity for the high (blue points) and low (violet points) gain. The two linear fits with the corresponding parameters are shown. The fit residuals are displayed in the middle panel. The $\pm$5\% limits are shown by the dashed gray lines. The bottom panel shows the ratio between the two gains. The linear function $y = ax + b$ has been used in \nimaAres{all three} cases, where $y$ refers to the quantity on the ordinate and $x$ to the quantity on the abscissa (i.e. the illumination charge.) \nimaA{The error bars are  statistical only.} } 
    \labfig{linearity}
\end{figure}

\subsection{Charge resolution}

The charge resolution represents the level of variability in measuring the number of photo-electrons detected by a PMT when exposed to light pulses of known intensity. It plays a crucial role in NectarCAM as it directly influences the accuracy of the final energy resolution for gamma-ray observations. The charge resolution of the Front-End Boards (FEBs) is calculated using the same measurements employed for the linearity test. For each measurement, the mean charge ($\overline{Q}$) and the standard deviation ($\sigma_Q$) are computed from the charge distribution of each pixel. The charge resolution is then obtained as $r = \sigma_Q/\overline{Q}$. \jonat{ The charge resolution for pulses $\overline{Q} > 400$ p.e. is measured using the LG channel. }

\reffig{charge_resolution} illustrates the distribution of charge resolutions measured with the 10 FEBv6 boards as a function of the mean charge in p.e. units. The charges measured using the LG channel are corrected for the $b_{\rm{HG}/\rm{LG}}$ ratio calculated in the linearity analysis. The solid gray line represents the statistical lower limit corresponding to Poisson statistics ($1/\sqrt{\overline{Q}}$). The charge resolution follows the Poissonian limit, showing that the main source of fluctuations is the number of photo-electrons. \nimaA{At the highest illumination, the resolution gets \nimaAres{worse} because of the underestimation of charges due to saturation of the anode signal in the PACTA LG channel \cite{2021NIMPA100765413T}.}

 

\begin{figure}[t!]
    \centering
    \includegraphics[width=\columnwidth]{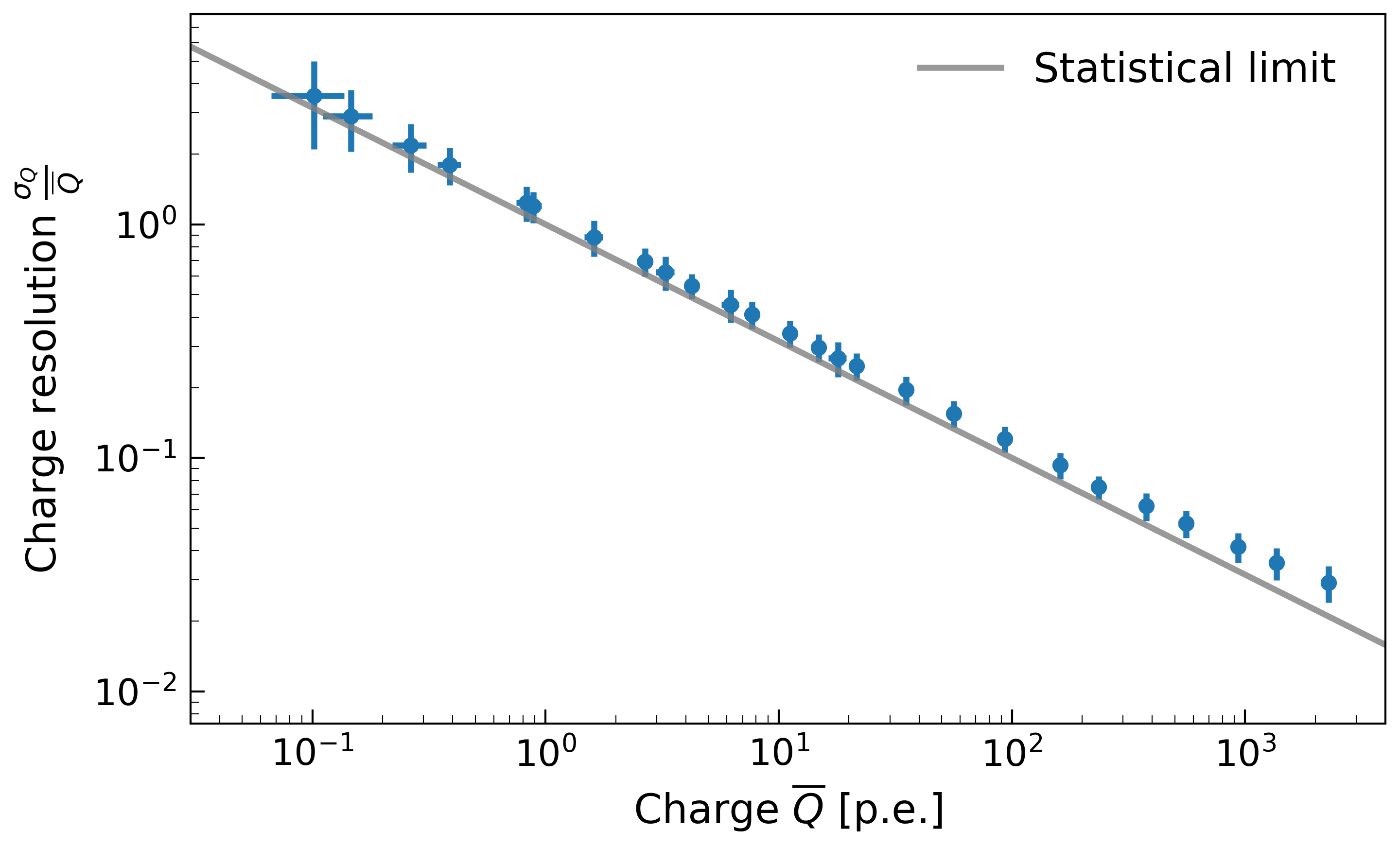}
    \caption{Charge resolution of 10 FEBv6 as a function of mean charge in p.e. measured with FFCLS and filters. The statistical Poisson lower limit is shown by the gray line.}
    \labfig{charge_resolution}
\end{figure}

\subsection{Single pixel timing precision}
The light's arrival time in each pixel is \nimaA{important} information that can be used to reduce the noise in shower images and improve the imaging cleaning and discrimination between Cherenkov photons and background. In this section, \sapoob{the averaged single pixel timing precision is estimated.} \nimaB{The camera focal plane is illuminated spatially uniformly with a laser source at a frequency of 1~kHz with \nimaA{pulse} energies between 8~pJ and 20~pJ.
The position of the maximum sample inside the 60~ns readout window (Time of Maximum, (TOM)) of each photon pulse is measured for each pixel. }
The TOM is estimated from the waveform after subtracting the pedestal. Two methods have been used. The \nimaA{first identifies} the position of the largest peak of the waveform using the function \texttt{signal.find\_peaks} from the \texttt{scipy} python package~\cite{scipy}. The \nimaA{second fits a Gaussian} to the largest peak of the waveform using as input the position of the peak from the first method \cite{timing_paper}.

\nimaB{The systematic timing uncertainty for each pixel is estimated by calculating the root mean square (RMS) of the obtained TOM distributions.} 
\reffig{pixel_res} displays the weighted mean of the RMS over all pixels as a function of the illumination charge, using both methods. The weight is determined by the inverse of the square of the standard deviation of the TOM distribution for each pixel. For incoming light intensities above approximately \nimaA{10~p.e.}, both methods indicate that the pixel precision is below \nimaA{1 ns; the Gaussian fit provides little improvement}. However, as shown by the solid gray line, \nimaA{above} an illumination charge of \nimaA{90~p.e.}, the precision \nimaA{is limited to the quantization noise of the trigger arriving randomly within the 1ns NECTAr3 samples, corresponding to an RMS of $1/\sqrt{12}~\mathrm{ns} = 290$ ps~\cite{timing_paper}.}
    
\begin{figure}[t!]
    \centering
    \includegraphics[width=\columnwidth]{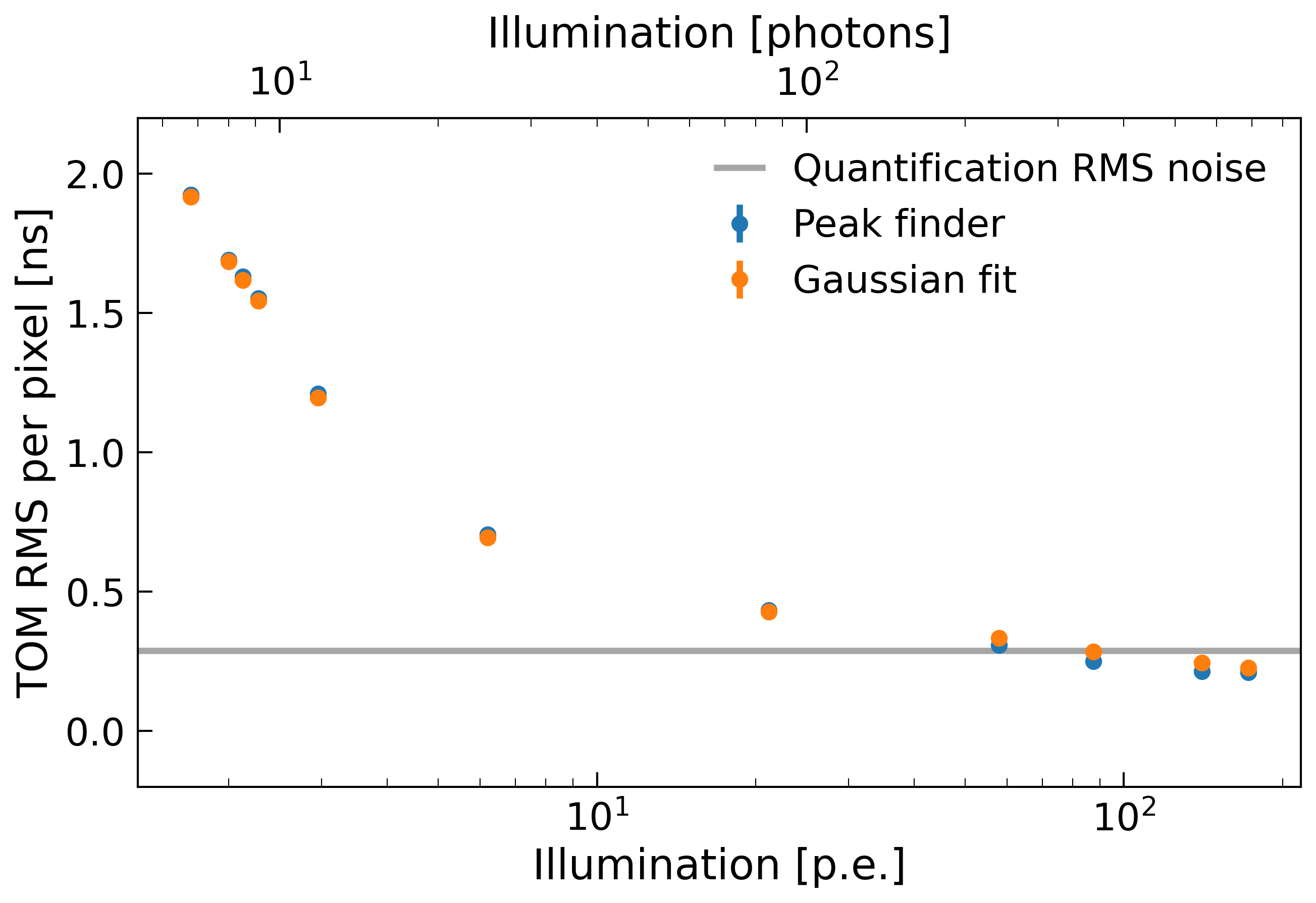}
    \caption{\sapoob{Single pixel timing precision averaged over all pixels} (in ns) \nimaA{as a function of photons (upper horizontal axis) and photoelectrons (lower horizontal axis)}. The timing resolution is given by the mean of the RMS distribution over all the 1855 pixels. Both methods are shown (in blue and orange). The gray solid line shows the quantization (RMS) noise given by $\frac{1}{\sqrt{12}}$~ns. }
    \labfig{pixel_res}
\end{figure}


\section{Conclusions}

This paper provides a comprehensive overview of the design and performance of the updated version of the FEB used in the CTAO's NectarCAM camera. The FEB is a crucial component responsible for sampling, digitizing, and transmitting the signals from the Cherenkov light captured by the PMTs in the camera. It also implements a local, pixel-level trigger and remote control of various components via SPI. 
The FEB \nimaA{is} a 12-layer printed-circuit-board with controlled impedance \nimaA{carrying} a Cyclone V INTEL FPGA and three different ASICs. The ASICs include the L0 trigger ASIC, four amplifiers for CTA (ACTA), and seven new \nectar{} ASICs, which are the focus of this paper.

The \nectar{} chip is \nimaA{the} crucial innovation in the new FEBv6, continuously sampling \nimaA{low and high gain} analog differential signals from the PMTs and utilizing a switched capacitor-based analog memory as a circular buffer until a trigger event occurs. Upon triggering, the relevant samples are read back and digitized by an on-chip ADC before transferring them to the FPGA. The \nectar{} chip's new feature, the ping-pong mode, significantly reduces the readout \nimaA{camera deadtime fraction from $5.2\%$ to $0.5\%$ at 7~kHz}.  \sapoob{This reduction could enable higher trigger rates and help lowering the trigger threshold to approximately 50 GeV at \nimaA{a} trigger rate of 7~kHz.}


Verification tests were conducted on the 10 new FEBv6 modules with the upgraded \nectar{} chips. Pixel timing precision measurements demonstrated precision below 1 ns for incident light intensities above approximately 20 photons. The linearity test of the FEBv6 readouts showed good linearity within 5\% deviation over a wide range of incident light intensities, spanning from 0.04 to 2000~p.e., resulting in a readout dynamic range greater than 3 decades. 

\section{Acknowledgements}

We thank the anonymous referees for carefully reading our manuscript and providing helpful suggestions that improved the clarity of the text. This work was conducted within the framework of the CTA Consortium. We are particularly thankful to the CTAO reviewers, Oscar Blanch and Richard White, for their diligent review and invaluable insights that have enriched our work. We gratefully acknowledge financial support from the following agencies and organizations: Ministry of Higher Education and Research, CNRS-INSU and CNRS-IN2P3, CEA-Irfu, ANR, Regional Council Ile de France, Labex ENIGMASS, OCEVU, OSUG2020 and P2IO, France; DESY, Helmholtz Association, Germany; Spanish Research State Agency (AEI) through the grant PID2019-104114RB-C32.


\appendix

\section{Saturation of the \nectar{} ADC}\labsec{saturat}

This appendix discusses the impact of saturation in the \nectar{} chip ADC and proposes mitigation strategies.  
The ADC in the \nectar{} chip consists of several MDAC stages (\reffig{pipeADC}), \nimaA{each responsible for amplifying signal denoted $V_{\rm ov}$, that can be out of range in case of saturation. The output voltage of each MDAC is given by the following equation.}

\begin{equation}
MDACo_{\rm i} = 2^{i} \cdot (V_{\rm ov} - C_{\rm Range}) + C_{\rm Range}
\end{equation}
Here, ${MDACo_{\rm i}}$ represents the output voltage of the MDAC of order $i$, and $C_{\rm Range} = 2 {\rm V}$ is the coding range of the ADC. As the signals progress through the final stages of the ADC, their voltages may increase significantly, potentially leading to deep saturation of the MDACs. This saturation can result in a recovery time required before the MDACs can accurately convert the next set of samples.

\begin{figure}[h]
    \centering
    \includegraphics[width=\columnwidth]{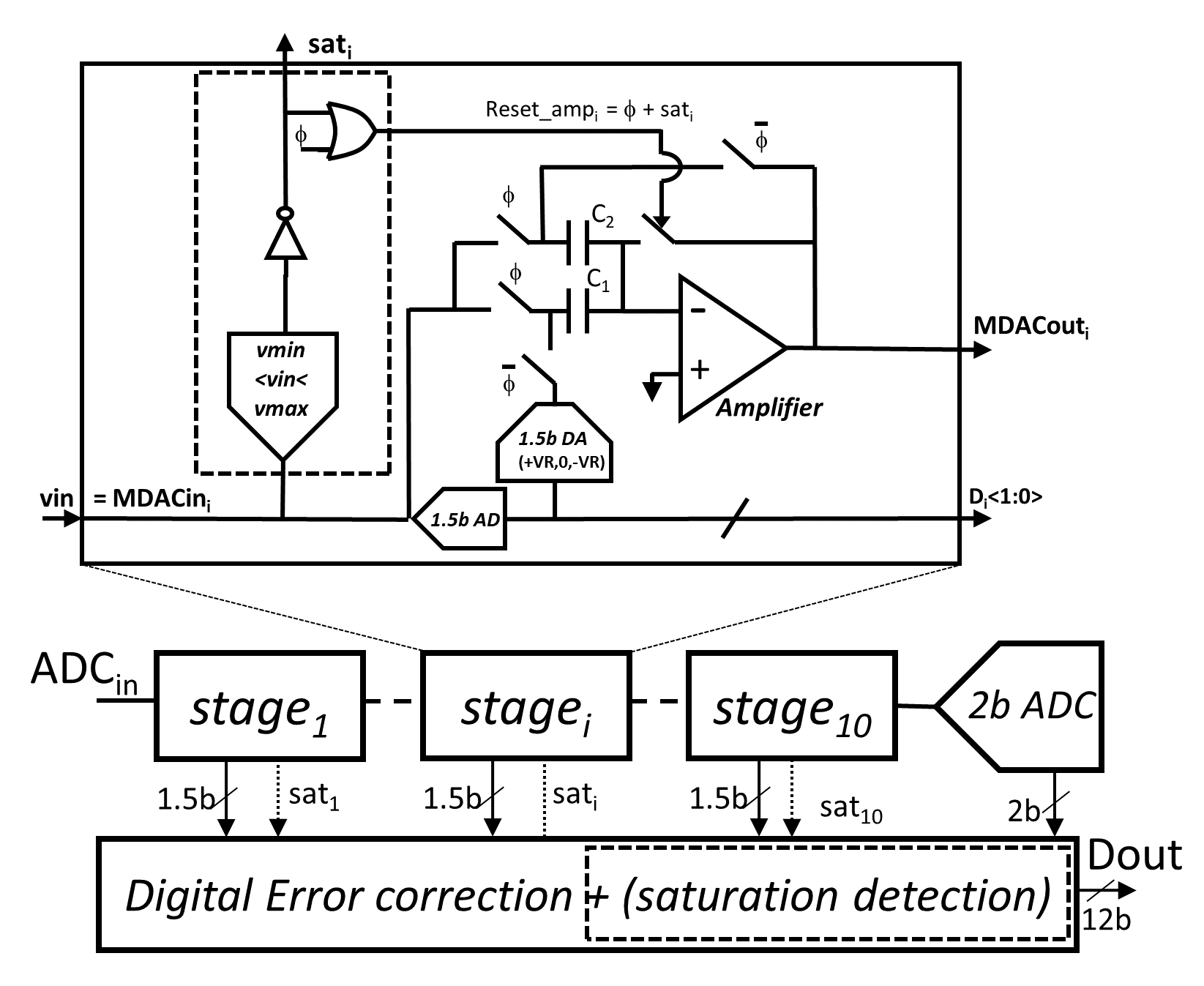}
    \caption{Simplified block diagram of the pipeline ADC integrated into the \nectar{} chip. The parts surrounded by a dotted line are those added to realize the anti-saturation system. $\protect\Phi$ and $\protect\bar{\Phi}$ are the two complementary clocks sequencing the ADC.}
    \labfig{pipeADC}
\end{figure}

In the configuration of the NectarCAM front-end board, both gains of the same PMT channel are processed by the same \nectar{} chip. Consequently, it is common for the conversion of a LG sample to follow immediately after an over-ranged HG sample. With the original version of NECTAr, this resulted in a degradation of the quality of the LG data, as demonstrated in \cite{TheseTsiahina}. 
The degradation in low gain data quality was mainly characterized by a significant increase in the differential non-linearity (DNL) and the appearance of missing codes, as depicted in \reffig{codedensity}. These issues, in turn, resulted in degraded integral nonlinearity (INL), reaching up to $\pm 1$\% deviation (\reffig{INL}). Additionally, noisy codes with increased noise levels, sometimes amplified by a factor of 10, were observed (\reffig{sdev}).


\begin{figure}[h]
    \centering
    \includegraphics[width=\columnwidth]{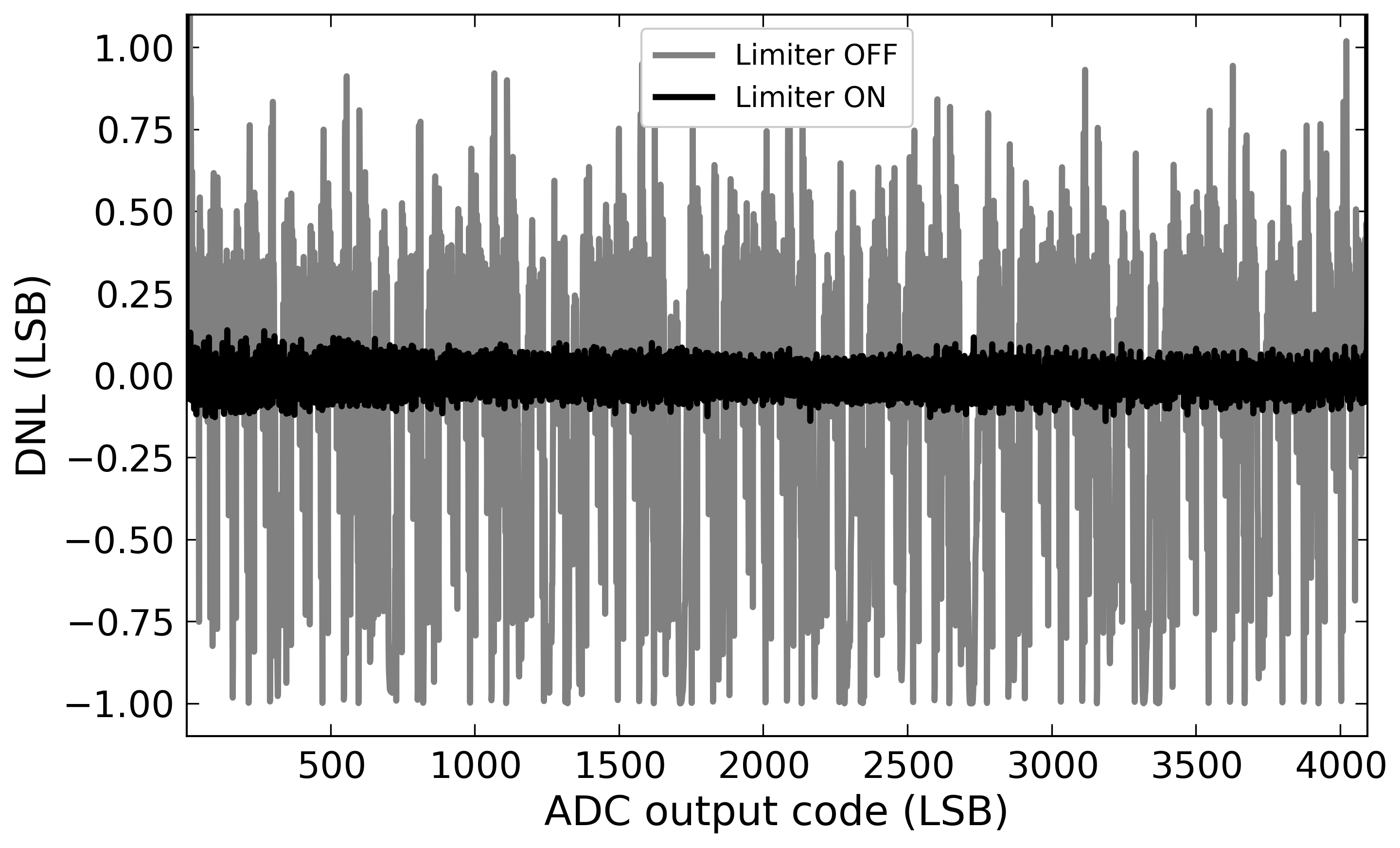}
    \caption{Densities of code for a LG channel when the HG channel exceeds the ADC range (2 V) by 0.65 V with limiter disabled (\nectar~) or enabled (\nectar{}). The differential non-linearity can be easily derived from these plots. With the limiter on, it is better than $\pm$ 0.15 LSB. These characteristics have been measured on the chip test bench using 16-bit DACs, with good DNL, \nimaAres{allowing one to} scan the input range of the chip. }
    \labfig{codedensity}
\end{figure}

In the upgraded H.E.S.S. camera \cite{HESSUpgrade2019}, which uses an early version of  the \nectar{} chip, clipping diodes were added to the inputs of the \nectar{} chips to mitigate this effect, at the cost of a 20\% loss in the dynamic range of the system. In \nectar{} , a window comparator has been added to each MDAC stage, as shown in \reffig{pipeADC}. It checks that its input signal is within a window corresponding to the ADC range extended by 15\%. If not, the MDAC amplifier is reset, thus preventing saturation of the next stages. Simultaneously, the digital output code of the ADC is forced to either the minimum or maximum code, depending on the bits already converted. This solution, based on well-proven comparator structures, avoids changes in the layout-sensitive \nimaA{analog} part of the ADC, thus minimising risks. 
\begin{figure}[h]
    \centering
    \includegraphics[width=\columnwidth]{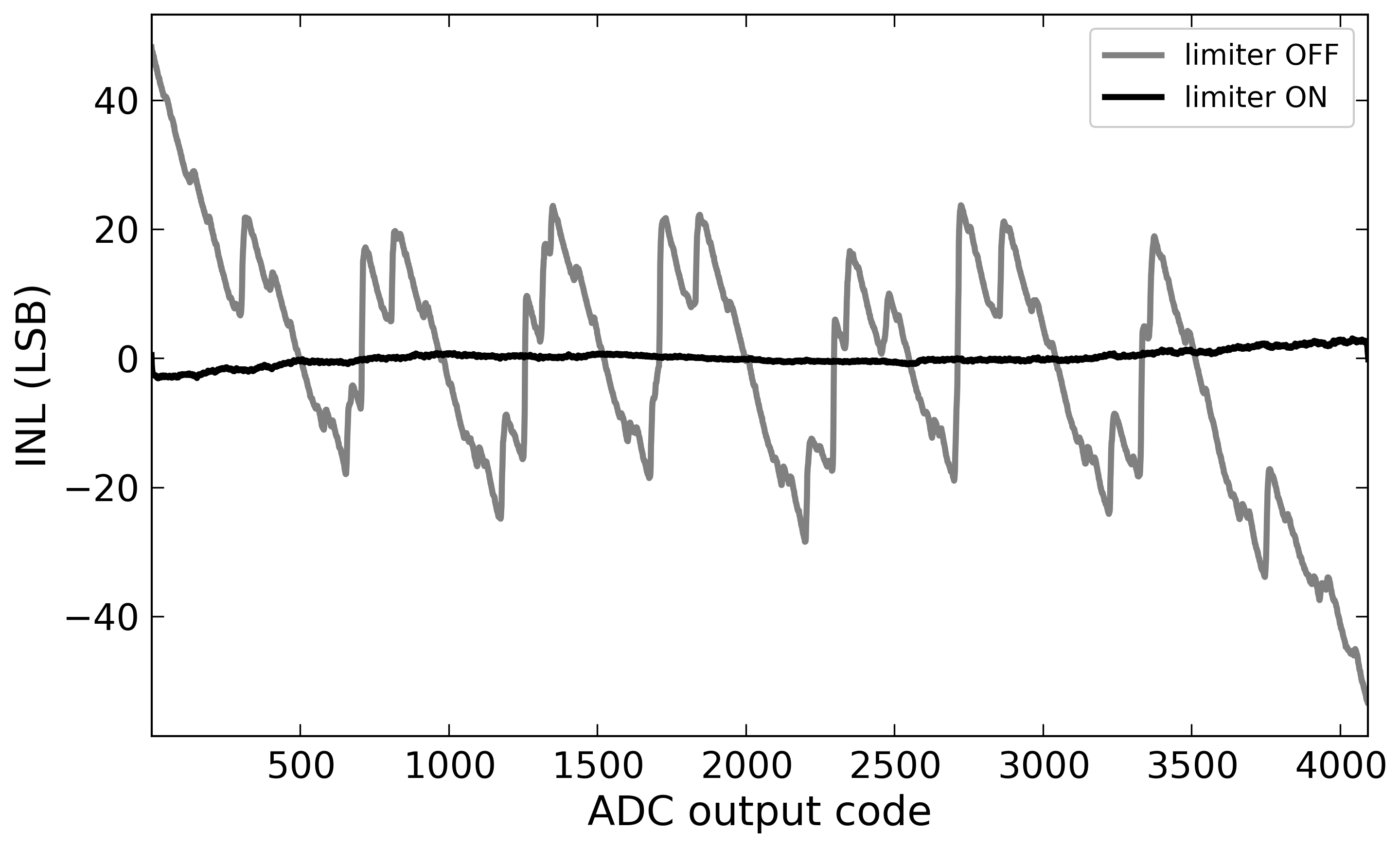}
    \caption{\nimaA{Differential non-linearity of a LG channel when the HG channel exceeds the ADC range (2~V) by 0.65~V with limiter disabled or enabled. These characteristics have been measured using the usual statistical method on the chip testbench using 16-bit DACs, with good DNL, allowing to scan the input range of the chip with a uniform voltage distribution. The DNL is drastically improved from $\pm1$~LSB to better than $\pm0.15$~LSB when the limiter is enabled.}}
    \labfig{INL}
\end{figure}

\begin{figure}[h]
    \centering
    \includegraphics[width=\columnwidth]{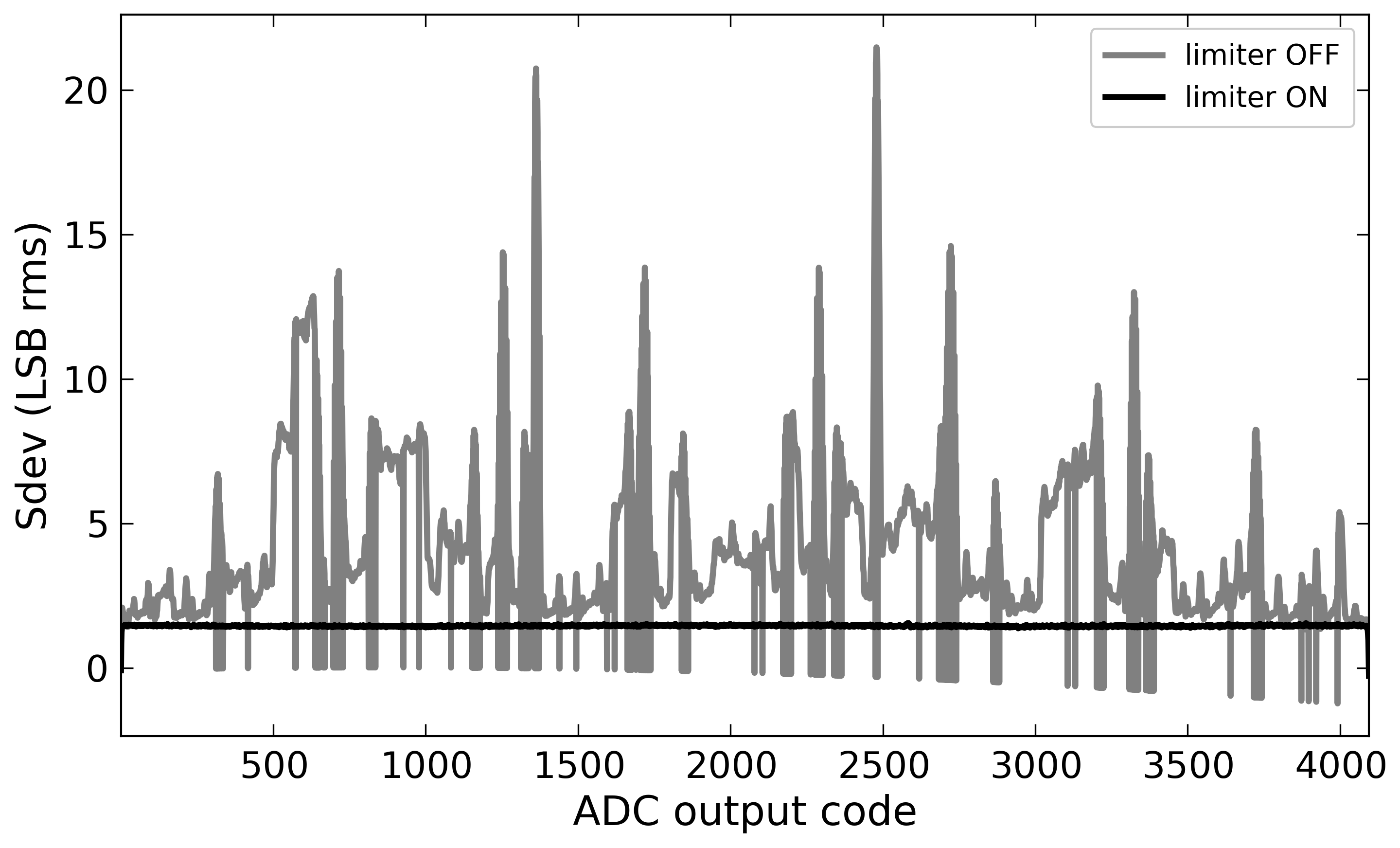}
    \caption{Measured standard deviation for LG channel in same conditions than for \reffig{codedensity}. With the limiter enabled, the standard deviation comes back to its nominal value (1.4 LSB rms).}
    \labfig{sdev}
\end{figure}

The effectiveness of this solution is highlighted by the gray curves in \reffig{codedensity}, \reffig{INL} and \reffig{sdev}: when the anti-saturation system is enabled on \nectar{}, the DNL, INL and noise characteristics on the low gain (with high gain over range) come back to their nominal values.

\section{Glossary}
\labapp{glossary}

\begin{description}
\item[ACTA:] Amplifier for the Cherenkov Telescope Array 
\item[ADC:] \nimaA{analog} to Digital Converter
\item[ASIC:] Application-Specific Integrated Circuit
\item[CTA:] Cherenkov Telescope Array
\item[CTAO:] Cherenkov Telescope Array Observatory
\item[DNL:] Differential Non-Linearity
\item[DTBP:] Digital Trigger BackPlane
\item[EPCQ64A:] \nimaA{Enhanced Programmable Configuration (device) Quad-Serial, Intel configuration PROM for FPGAs}
\item[FEB:] Front End Board
\item[FFCLS:] Flat Field Calibration Light Source 
\item[FIFO:] First-In-First-Out
\item[FPGA:] Field-Programmable Gate Array
\item[FPM:] Focal Plane Module
\item[FSM:] Finite State Machine
\item[GMII:] Gigabit Media-Independent Interface
\item[GSPS:] Giga Sample Per Second
\item[HG:] High Gain
\item[HVPA:] High Voltage and Pre-Amplification board 
\item[IB:] Interface Board
\item[IACT:] Imaging Atmospheric Cherenkov Telescope
\item[INL:] Integral Non-Linearity
\item[IP:] Internet Protocol
\item[JTAG:] Joint Test Action Group electronic bus
\item[L0:] Level 0 trigger (pixel level)
\item[L1:] Level 1 trigger (camera level)
\item[LG:] Low Gain
\item[LSB:] Least Significant Bit
\item[LST:] Large Sized Telescope
\item[MAC:] Media Access Control address
\item[MC:] Monte-Carlo simulations
\item[MDAC:] Multiplying Digital-to-Analog Converter
\item[MSPS:] Mega Sample per second
\item[MST:] Medium Sized Telescope of CTA
\item[NMC:] NectarCAM Module Controller
\item[NSB:] Night Sky Background
\item[OPCUA:] Open Platform Communications Unified Architecture
\item{PACTA}PreAmplifier for the CTA cameras
\item[PCB:] Printed Circuit Board
\item[PMT:] PhotoMultiplier Tube
\item[PROM:]Programmable Read Only Memory
\item[QFN:]Quad Flat no-leads package
\item[QFP:]Quad Flat Package
\item[RAM:] Random Access Memory   
\item[RBB:] Read Bottom Bus
\item[RMS:] Root Mean Square
\item[SPI:] Serial Peripheral Interface
\item[SST:] Small Sized Telescope of CTA
\item[UDP:] User Datagram Protocol
\item[VHE:] Very-High Energy
\item[XML:] Extensible Markup Language
\end{description}

\bibliographystyle{elsarticle-num}
\bibliography{mybibfile}

\end{document}